\documentclass{aastex62}

\usepackage{xspace}

\newcommand\msun{\ensuremath{\mathrm{M}_\odot}\xspace}

\received{}
\revised{}
\accepted{9-21-2021}

\shorttitle{Small disks in Orion}
\shortauthors{Otter and Ginsburg et. al.}

\usepackage{savesym}
\savesymbol{tablenum}
\usepackage{siunitx}
\usepackage{pbox}
\restoresymbol{SIX}{tablenum}

\begin{document}

\title{Small Protoplanetary Disks in the Orion Nebula Cluster and OMC1 with ALMA}

\correspondingauthor{Justin Otter}
\author[0000-0003-3191-9039]{Justin Otter}
\email{jotter2@jhu.edu}
\affil{REU Student at the National Radio Astronomy Observatory, 1003 Lopezville Rd, Socorro, NM 87801 USA}
\affil{Department of Physics \& Astronomy, Johns Hopkins University, Bloomberg Center, 3400 N. Charles St, Baltimore, MD 21218, USA}

\author[0000-0001-6431-9633]{Adam Ginsburg}
\affil{Jansky fellow of the National Radio Astronomy Observatory, 1003 Lopezville Rd, Socorro, NM 87801 USA}
\affil{Department of Astronomy, University of Florida, PO Box 112055, USA}

\author[0000-0002-4276-3730]{Nicholas P. Ballering}
\affil{Department of Astronomy, University of Virginia, Charlottesville, VA 22904, USA }

\author[0000-0001-8135-6612]{John Bally}
\affil{Center for Astrophysics and Space Astronomy, Department of Astrophysical and Planetary Sciences, University of Colorado, Boulder}

\author{J.A. Eisner}
\affil{Steward Observatory, University of Arizona, 933 N Cherry Avenue, Tucson, AZ 85721 USA}

\author[0000-0002-2542-7743]{Ciriaco Goddi}
\affil{Department of Astrophysics/IMAPP, Radboud University Nijmegen, P.O. Box 9010, NL-6500 GL Nijmegen, The Netherlands}
\affil{ALLEGRO/Leiden Observatory, Leiden University, P.O. Box 9513, NL-2300 RA Leiden, The Netherlands}

\author[0000-0001-6765-9609]{Richard Plambeck}
\affil{Radio Astronomy Lab, University of California, 501 Campbell Hall, Berkeley, CA 94720-3441, USA}

\author[0000-0002-9154-2440]{Melvyn Wright}
\affil{Radio Astronomy Lab, University of California, 501 Campbell Hall, Berkeley, CA 94720-3441, USA}

\begin{abstract}
The Orion Nebula Cluster (ONC) is the nearest dense star-forming region at $\sim$400 pc away, making it an ideal target to study the impact of high stellar density and proximity to massive stars (the Trapezium) on protoplanetary disk evolution.
The OMC1 molecular cloud is a region of high extinction situated behind the Trapezium in which actively forming stars are shielded from the Trapezium's strong radiation. 
In this work, we survey disks at high resolution with ALMA at three wavelengths with resolutions of 0.095\arcsec\xspace (3 mm; Band 3), 0.048\arcsec\xspace (1.3 mm; Band 6), and 0.030\arcsec\xspace (0.85 mm; Band 7) centered on radio Source I.
We detect 127 sources, including 15 new sources that have not previously been detected at any wavelength.
72 sources are spatially resolved at 3 mm, with sizes from $\sim$8 - 100 AU.
We classify 76 infrared-detected sources as foreground ONC disks and the remainder as embedded OMC1 disks.
The two samples have similar disk sizes, but the OMC1 sources have a dense and centrally concentrated spatial distribution, indicating they may constitute a spatially distinct subcluster.
We find smaller disk sizes and a lack of large ($>75$ AU) disks in both our samples compared to other nearby star-forming regions, indicating that environmental disk truncation processes are significant.
While photoevaporation from nearby massive Trapezium stars may account for the smaller disks in the ONC, the embedded sources in OMC1 are hidden from this radiation and thus must truncated by some other mechanism, possibly dynamical truncation or accretion-driven contraction.
\end{abstract}


\defcitealias{andrews_scaling_2018}{A18}
\defcitealias{eisner_protoplanetary_2018}{E18}
\defcitealias{muench_luminosity_2002}{MLLA}

\defcitealias{robberto_wide-field_2010}{[1]}
\defcitealias{chambers_pan-starrs1_2016}{[2]}
\defcitealias{friedel_high_2011}{[3]}
\defcitealias{robberto_hubble_2013}{[4]}
\defcitealias{sheehan_vla_2016}{[5]}
\defcitealias{shuping_new_2004}{[6]}
\defcitealias{forbrich_population_2016}{[7]}
\defcitealias{feigelson_x-ray-emitting_2002}{[8]}
\defcitealias{nissen_observations_2007}{[9]}
\defcitealias{chen_bandmerged_2016}{[10]}
\defcitealias{odell_nature_2015}{[11]}

\section{Introduction} \label{sec:intro}

Protoplanetary disks play important roles both in star and planet formation. The lifetime and properties of disks impact the evolution of the central protostar because disks act as gas reservoirs from which protostars can continue to accrete. 
Because disks are progenitors of planetary systems, measurements of disk sizes and masses provide constraints on the mass reservoir available for planet formation \citep[e.g.][]{mordasini_global_2015}, while disk lifetimes can constrain giant planet formation models \citep[e.g.][]{alibert_models_2005, winn_occurrence_2015}.

The nearest star-forming clouds containing protoplanetary disks have a low stellar density (such as Taurus, Lupus, Upper Sco, etc.), which has meant that much of our understanding of disk evolution has focused on relatively isolated disks \citep[e.g.][]{andrews_circumstellar_2005, ansdell_alma_2016, tazzari_physical_2017, barenfeld_alma_2016}.
However, the majority of stars form in massive, rich clusters, so to understand the most common mode of disk evolution in our Galaxy we must probe these dense clusters \citep{carpenter_2mass_2000, lada_infrared_1991, lada_environments_1993, lada_embedded_2003}.
Observations and simulations show that the cluster environment can have significant impacts on disk properties and evolution \citep[e.g.][]{breslau_sizes_2014,guarcello_photoevaporation_2016}.
Both disk dust mass and radius are directly observable properties that are sensitive to the environment \citep{parker_external_2021}.

Disks are shaped by two key environmentally-dependent processes: external photoevaporation and dynamical truncation.
Disks in regions with nearby massive stars are externally eroded by the intense ionizing radiation field.
External photoevaporation therefore plays a strong role in driving disk truncation in O-star-containing regions.
This effect has been long observed in the Orion Nebula Cluster (ONC) from the cometary morphology of many of the protoplanetary disks, where the dusty tails have been shown to point away from luminous Trapezium stars \citep[the `proplyds'][]{odell_discovery_1993, mccullough_photoevaporating_1995, bally_externally_1998}.
Dynamical truncation impacts disks primarily through disk-disk and disk-star interactions (fly-bys), which scale with stellar density \citep[e.g.][]{vincke_cluster_2016}.
Simulations from \citet{winter_protoplanetary_2018} suggest that external photoevaporation dominates disk evolution over dynamical interactions, as clusters with a high enough density for dynamical interactions to be significant also tend to contain massive stars with strong UV fields.

The ONC is a gas-rich star-forming region, and is the nearest site of high mass star formation.
Recent distance measurements to the ONC range from 388 $\pm$ 5 pc \citep{kounkel_goulds_2017} to 414 $\pm$ 7 pc \citep{menten_distance_2007}; we adopt  $d=400$ pc as an easily-rescaled compromise.
The proximity of the ONC allows us to study the disk population in detail; we can detect fainter disks and resolve the larger disks.
The ONC contains main-sequence stars and exposed protostars with stellar ages $\lesssim$2 Myr centered on the Trapezium \citep[e.g.][]{jeffries_no_2011, fang_improved_2020}.
Stars in the ONC are subject to only a couple magnitudes of extinction at optical wavelengths \citep[e.g.][]{johnson_infrared_1967, odell_high_2000}, and many are surrounded by disks seen in silhouette or rendered visible by ionized gas  \citep[e.g., the proplyds;][]{storzer_photodissociation_1999, eisner_proplyds_2008}.

The OMC1 cloud core, located $\sim$90\arcsec\xspace northwest of and few tenths of a parsec behind the Trapezium, is younger than the ONC and is actively forming stars \citep{zuckerman_model_1973, genzel_orion_1989, bally_overview_2008}.
At the core of OMC1 is the Becklin-Neugebauer/Kleinmann-Low (BN-KL) region, which contains two high mass stars: BN \citep[$\sim10M_\odot$][]{goddi_multi-epoch_2011} and Src I \citep[$M\approx15M_\odot$][]{ginsburg_keplerian_2018}.
The warm, dense material surrounding Src I is known as the `hot core'.
Stars in OMC1 are highly embedded, with a mean extinction of  $A_V\sim30$ magnitudes, while ONC stars have extinctions of only a few magnitudes \citep{scandariato_extinction_2011}.
The large contrast in ONC and OMC1 extinction implies that the dust rendering OMC1 sources optically invisible also shields the disks from the ionizing radiation from Trapezium stars in the ONC, so despite the proximity of OMC1 to the ONC, they occupy different environments from a disk-evolution perspective.

High dust extinction in OMC1 obscures the region in the optical and infrared, so long-wavelength observations are needed to detect the deeply embedded disk-bearing young stars.
Millimeter wavelength radiation from young stellar objects (YSOs) is due primarily to dust in their circumstellar disks \citep[e.g.][]{hildebrand_determination_1983, beckwith_survey_1990}.
The intensity of radiation from dust depends on the optical depth: as the dust column density, and in turn the dust optical depth, increases, the emission spectrum approaches that of a black body.
The dust emission spectrum on the Rayleigh-Jeans tail is characterized by a power-law slope $\alpha$ known as the spectral index and defined as:
\begin{equation} \label{eq:alpha}
    S_\nu \propto \nu^\alpha 
\end{equation}
where $S_\nu$ is the intensity at frequency $\nu$, and $\alpha$ is the spectral index.
Optically thin dust exhibits a slope steeper than $\alpha > 2$ \citep{hildebrand_determination_1983},
while optically thick dust in the Rayleigh-Jeans regime exhibits $\alpha = 2$.
\citet{ballering_protoplanetary_2019} show that all but the most massive disks are expected to be only partially optically thick, with $\alpha \gtrsim 2.4$, while their optically thin disk model has $2.6 \leq \alpha \leq 3.0$ from 1-3 mm.

In this work, we use ALMA band 3, 6, and 7 (3, 1.3, and 0.85 mm) observations centered on Src I to measure the fluxes and sizes of protoplanetary disks to characterize and better understand the Orion YSO population in the ONC and OMC1.
In Section~\ref{sec:obs} we present the ALMA observations.
In Section~\ref{sec:analysis} we discuss source identification, cross-matching to other catalogs, disk fitting and measurement methods, and present the measured sizes, inclinations, and spectral indices of disks in the region.
In Section~\ref{sec:oncvsomc1} we compare the physical properties and spatial distributions of the ONC and OMC1 samples
In Section~\ref{sec:source_pop} we consider the radius-luminosity scaling relationship, and compare our disk sizes and masses to those in other regions.
Lastly, in Section~\ref{sec:discussion}, we discuss different disk truncation mechanisms in the context of our results in Orion.

\section{Observations and Methods} \label{sec:obs}

This work presents single pointing ALMA continuum observations of the Orion Nebula Cluster in bands 3, 6, and 7, centered on Src I (ICRS 05:35:14.51 -05:22:30.56).
Table~\ref{tab:bands} lists observational and imaging parameters for each image.
These observations have previously been presented in \citet{ginsburg_keplerian_2018}, who give further details on the data and calibration.

\begin{table}
\centering
\begin{tabular}{c|p{0.085\textwidth}|p{0.065\textwidth}|p{0.11\textwidth}|p{0.05\textwidth}|p{0.11\textwidth}|p{0.11\textwidth}|c|p{0.088\textwidth}|p{0.084\textwidth}}
Band & \pbox{0.085\textwidth}{Central Frequency} & \pbox{0.065\textwidth}{Primary Beam FWHM}  & \pbox{0.11\textwidth}{Beam \\ Major$\times$Minor} & \pbox{0.05\textwidth}{Beam PA} & \pbox{0.11\textwidth}{RMS \\ (central-edge)} & \pbox{0.11\textwidth}{Min. $M_{\text{disk}}$ $^a$ \\ (central-edge)} & \pbox{0.052\textwidth}{Robust} & \pbox{0.088\textwidth}{Flux Threshold$^b$} & \pbox{0.084\textwidth}{Baseline Lengths} \\ 
 & GHz & \arcsec & \arcsec $\times$ \arcsec & $^\circ$ & mJy beam$^-1$ & $M_\oplus$ & & mJy beam$^-1$ & \pbox{0.084\textwidth}{m \\ (k$\lambda$)} \\
\hline
3    &   98.0 & 59.4 & 0.097 $\times$ 0.071 & 43.4 & 0.04 - 1.0 & 2.3 - 57 &  0.5 & 0.05 & \pbox{0.084\textwidth}{21-14854 \\ (6.9-4855) \\} \\
6    &   223.5 & 26.1 & 0.049 $\times$ 0.037 & 80.4 & 0.2 - 2 & 5 - 50 & 0.5 & 1 & \pbox{0.084\textwidth}{150-12147 \\ (112-9055) \\} \\
7    &   339.8 & 17.1 & 0.030 $\times$ 0.024 & -69.6 & 0.12 - 0.45 & 2.0 - 7.4 & 0.5 & 0.05 & \pbox{0.084\textwidth}{210-1124 \\ (238-1273)} \\
\end{tabular}
\caption{Imaging parameters and information for each image. $^a$ Minimum detectable disk dust mass, calculated according to Equation~\ref{eq:dmass}, see Section~\ref{sec:dmasses}. $^b$ the flux level in the residual image to end \texttt{tclean}. \label{tab:bands}}
\end{table}

These data were calibrated and imaged in CASA \citep{mcmullin_casa_2007}, a software package designed for reducing interferometric data.
We initially imaged each data set with a different robust values (-2, 0.5, and 2), a variety of minimum flux thresholds for \texttt{tclean}, and baseline ranges.
We visually examined each image and compared flux measurements for different sources (measurement described in Section~\ref{sec:analysis}).
We discarded images with clear imaging artifacts that could impact source fitting.
Images with robust parameters of -2 yielded smaller disk sizes, as some of the extended disk flux was effectively resolved out, so we excluded these.
Band 6 and 7 images with all baselines included had disks in the central region with flux contamination from the extended `hot core' emission and were more affected by imaging artifacts than images with short baselines excluded, so we created images excluding visibilities shorter than 150 and 210 m for band 6 and 7 respectively.
These baseline cuts were the minimum cut needed to visually remove the emission contamination, and correspond to removing emission on scales greater than 1.8\arcsec\xspace and 0.85\arcsec\xspace respectively\footnote{The largest disk we detect (source 15) has an angular size of 0.22\arcsec\xspace and 0.24\arcsec\xspace in bands 6 and 7 respectively.}.
The uv coverage is similar in each band even with these cuts.
Each image is primary beam corrected.
We image the band 3 data to the 5\% primary beam recovery region, while we only the band 6 and 7 data to the 30\% recovery regions because the noise is too high to detect sources past 30\%.

Figure~\ref{fig:gemini} shows a Gemini image of OMC1  and the surrounding area \citep{bally_orion_2015} with our image fields-of-view shown in blue dashed lines.
We circle the mm continuum source locations identified at 3 mm (green), 1.3 mm (pink), and 0.85 mm (red).
Sources detected in multiple bands are circled with the color corresponding to the shortest wavelength detection.

\section{Analysis}\label{sec:analysis}

\subsection{Source Identification}

We identified compact sources in these images by eye.
Initially we employed the python package \texttt{astrodendro}\footnote{\url{https://dendrograms.readthedocs.io/en/stable/}}, which identifies sources with dendrograms.
While this process was useful in identifying some faint compact sources, this package also made a large number of spurious identifications of extended emission and imaging artifacts, and missed a small number of compact sources.
We begin with the \texttt{astrodendro} catalog for the band 3 image, remove spurious sources, and visually search the images for remaining sources. 
To ensure we do not miss any sources, we check the source locations of other studies of the region, including \citet{eisner_protoplanetary_2018} (hereafter \citetalias{eisner_protoplanetary_2018}), an ALMA band 7 study with some overlap, and the IR catalog \citet{muench_luminosity_2002} (hereafter \citetalias{muench_luminosity_2002}).
Finally, we require that each source has a peak signal-to-noise ratio S/N$\geq5$ in one or more bands to remain in our final catalog.
We estimate the noise by computing the RMS in an annulus around each source (see Section~\ref{sec:fitting} for more details).
For the band 6 and 7 images, we use the same source locations as band 3 (after visually checking for sources undetected in band 3, and find none).
We require a less stringent S/N $\geq$ 3 to include a band 6 or 7 source in our catalog because these sources also have band 3 detections, thus they are less likely to be spurious. 42/50 band 6 30/34 band 7 sources have S/N $>$ 5.

\begin{figure}
    \centering
    \includegraphics[width=\textwidth]{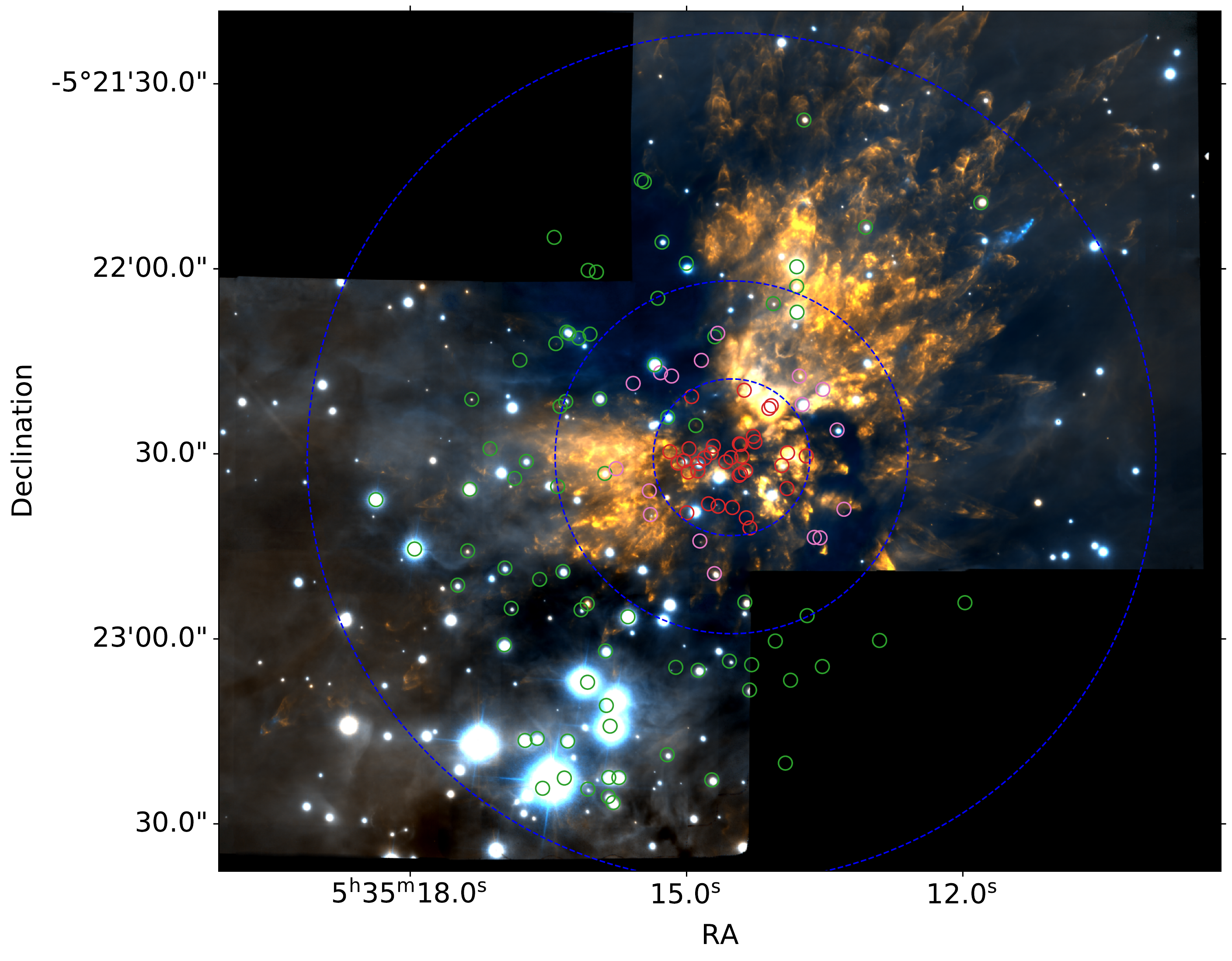}
    \caption{An IR image mosaic from Gemini of the ONC, with our detected sources circled \citep{bally_orion_2015}. The dotted shapes show the fields of view of our data in different bands, 3 mm, 1.3 mm, and 0.85 mm in descending size. Sources circled in red are detected at 0.85 mm, pink at 1.3 mm, and green at 3 mm.}
    \label{fig:gemini}
\end{figure}

\bigskip 
\subsection{Source Detections}
\begin{figure}[!h]
    \centering
    \includegraphics[width=0.7\textwidth]{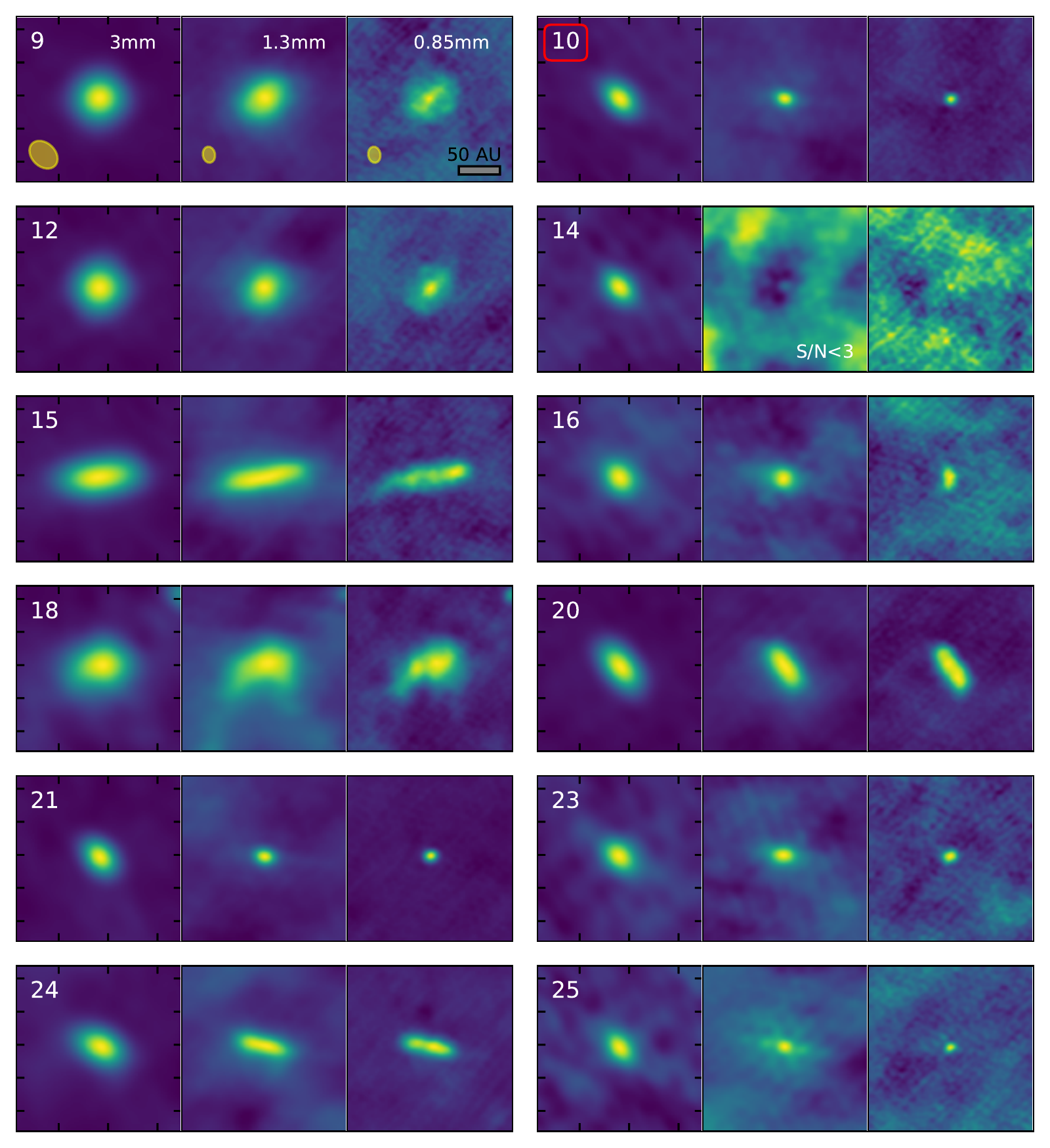}
    \caption{(1/3) Images of sources in the band 7 (0.85 mm) field of view, from left to right: band 3 (3 mm), band 6 (1.3 mm), and band 7. The yellow ellipses shows the synthesized beam size for each image. Each cutout is 0.5\arcsec\xspace across (200 AU). The number in the top left is the source ID, and source numbers boxed in red correspond to newly detected sources (see Section~\ref{sec:new_dets}). Nondetections are shown as well.} 
    \label{fig:b7_pg1}
\end{figure}

In total, we detect 127 sources in our band 3 (3 mm) data.
In our band 7 (0.85 mm) data, we detect 34/36 band 3 sources in the band 7 field of view, and in our band 6 (1.3 mm) data we detect 50/59 band 3 sources in the band 6 field of view.
Figure~\ref{fig:b7_pg1} highlights sources in the band 7 FOV, Figure~\ref{fig:b6_pg1} shows sources in the band 6 FOV (excluding band 7 sources), and Figure~\ref{fig:b3_pg1} shows all the remaining sources only in the band 3 FOV.
Appendix~\ref{app:zoom_plots} includes figures showing the locations of all detected sources.
\bigskip

\setcounter{figure}{1}
\begin{figure}[!h]
    \centering
    \includegraphics[width=0.7\textwidth]{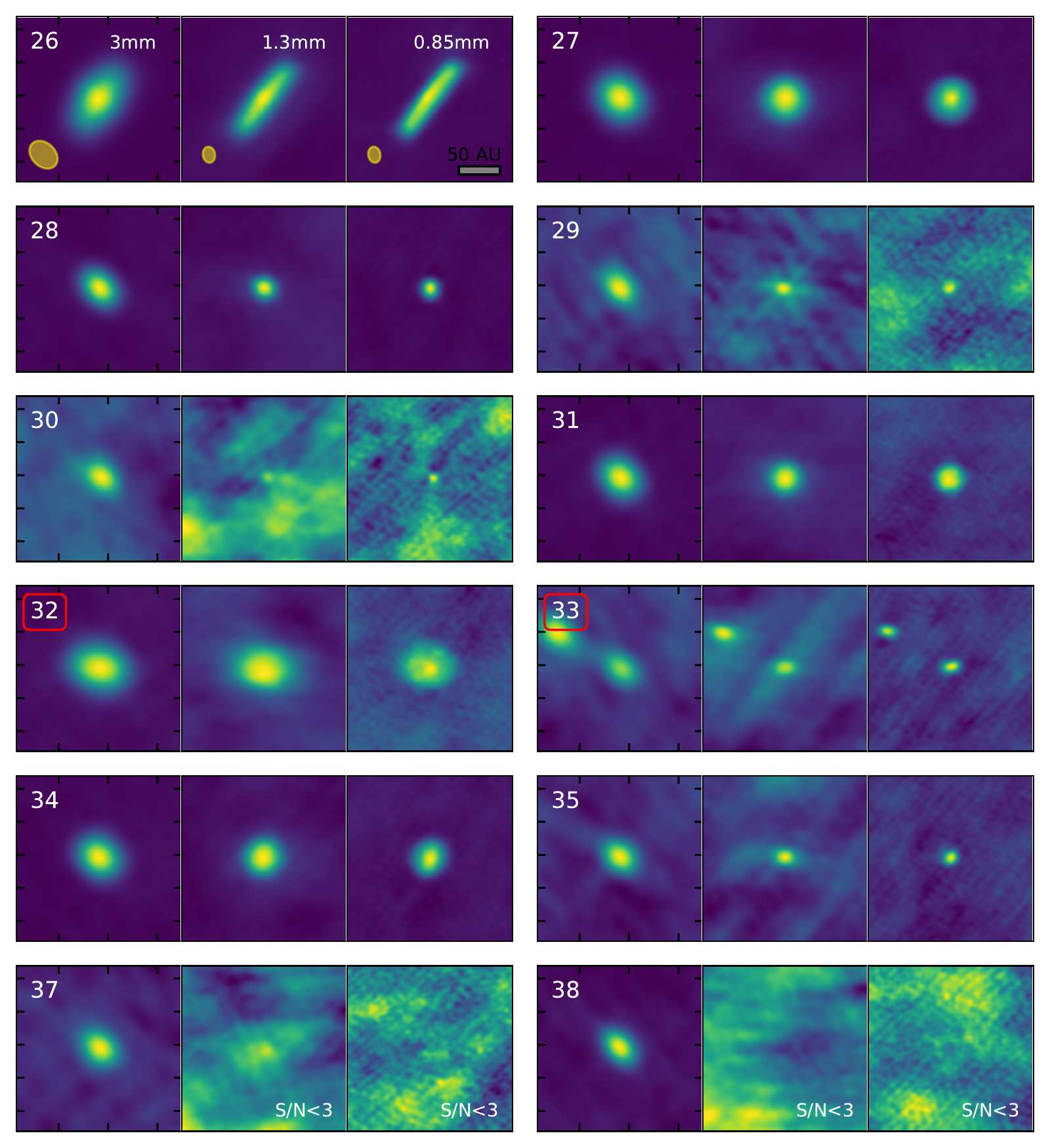}
    \caption{(2/3) Images of sources in the band 7 (0.85 mm) field of view, from left to right: band 3 (3 mm), band 6 (1.3 mm), and band 7. The yellow ellipses shows the synthesized beam size for each image. Each cutout is 0.5\arcsec\xspace across (200 AU). The number in the top left is the source ID, and source numbers boxed in red correspond to newly detected sources. Nondetections are shown as well.}
    \label{fig:b7_pg2}
\end{figure}
\setcounter{figure}{1}
\begin{figure}[!h] 
    \centering
    \includegraphics[width=0.7\textwidth]{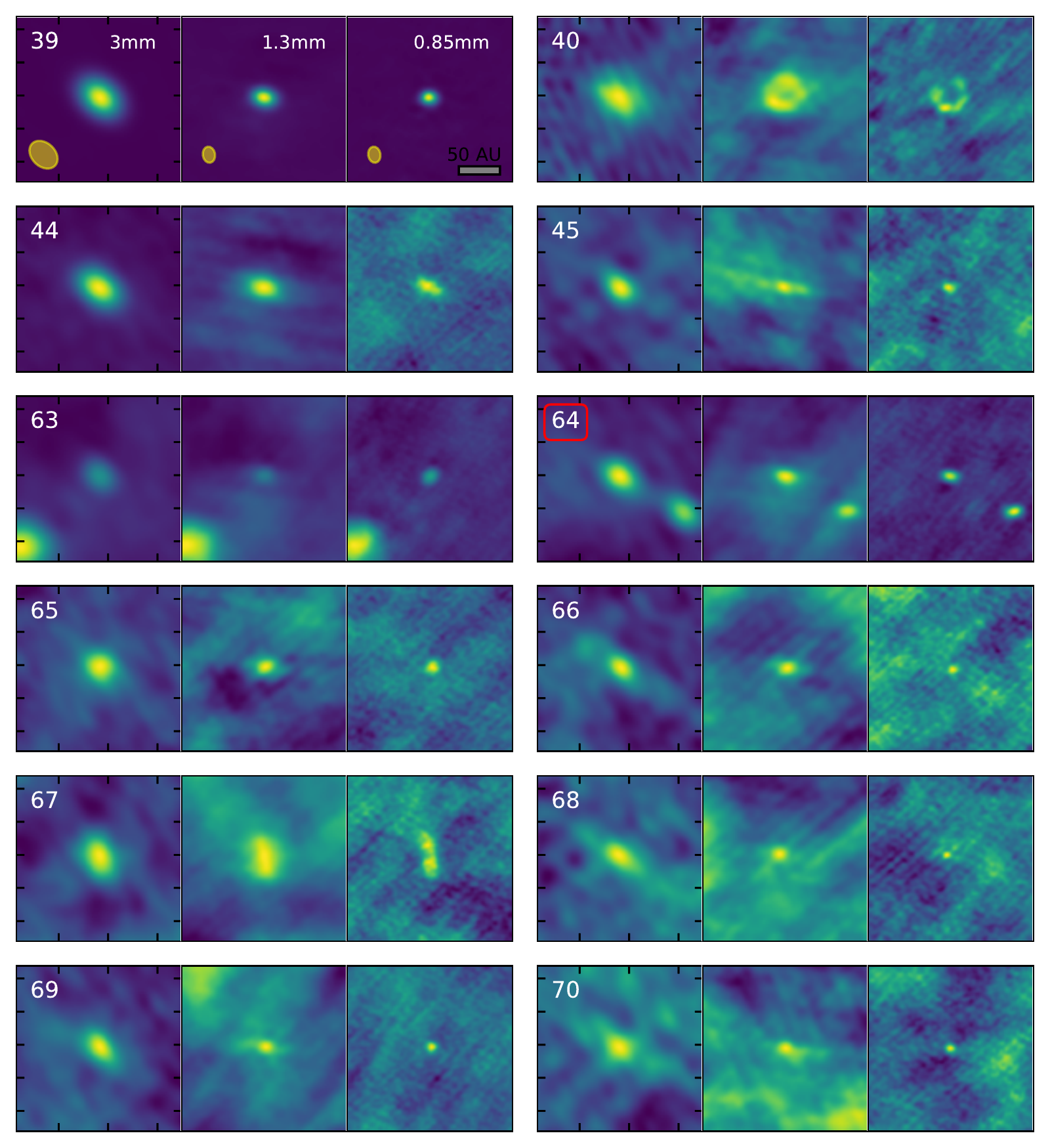}
    \caption{(3/3) Images of sources in the band 7 (0.85 mm) field of view, from left to right: band 3 (3 mm), band 6 (1.3 mm), and band 7. The yellow ellipses shows the synthesized beam size for each image. Each cutout is 0.5\arcsec\xspace across (200 AU). The number in the top left is the source ID, and source numbers boxed in red correspond to newly detected sources. Nondetections are shown as well.} 
    \label{fig:b7_pg3}
\end{figure}

\begin{figure}[!h] 
    \centering
    \includegraphics[width=0.7\textwidth]{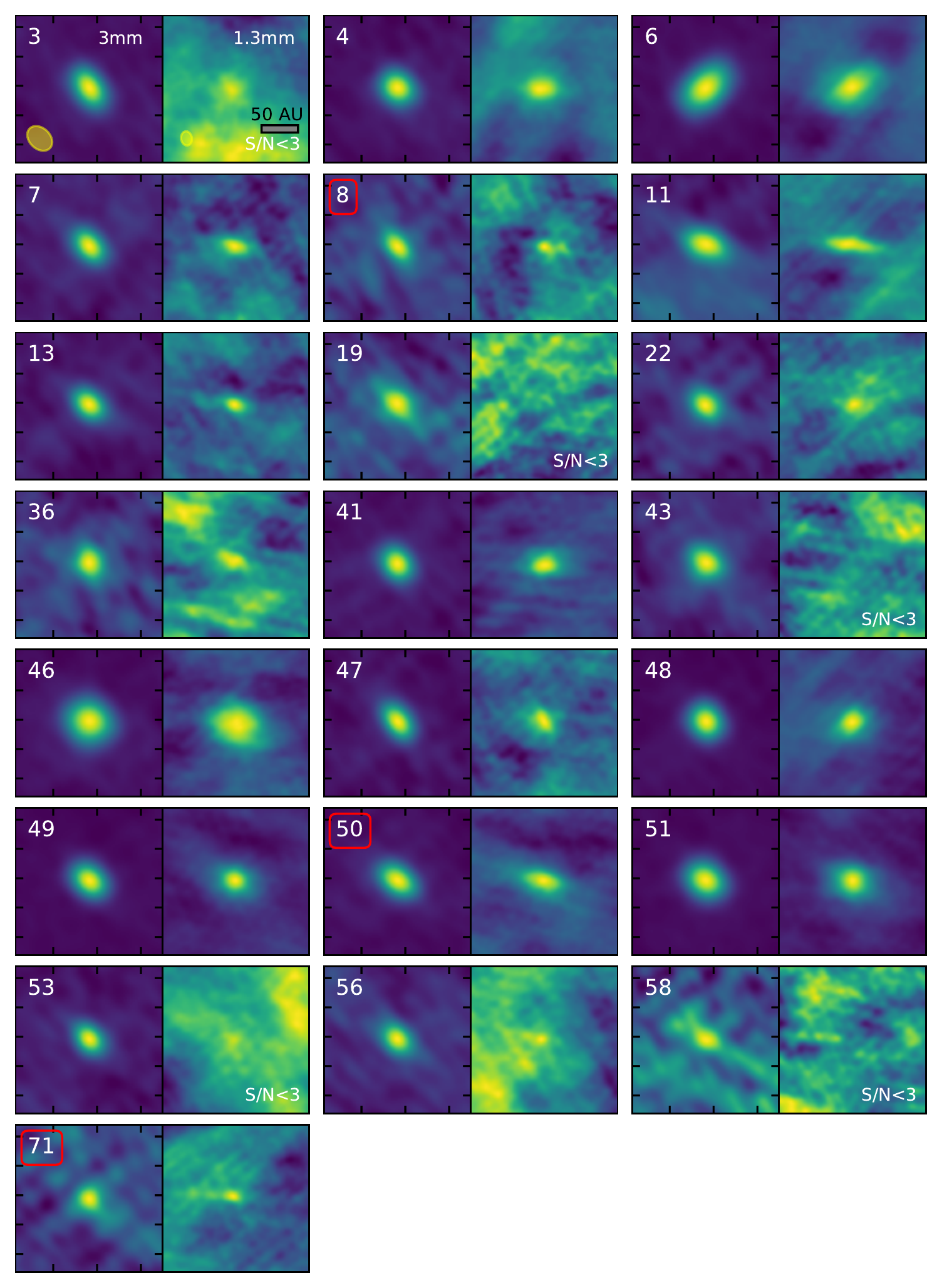}
    \caption{Images of sources detected in the band 6 (right panel) and band 3 (left panel) FOVs. The yellow ellipses shows the synthesized beam sizes. Each cutout is 0.5\arcsec\xspace across (200 AU). The number in the top left is the source ID, and source numbers boxed in red correspond to newly detected sources. Nondetections are shown as well.} 
    \label{fig:b6_pg1}
\end{figure}

\begin{figure}[!h] 
    \centering
    \includegraphics[width=0.7\textwidth]{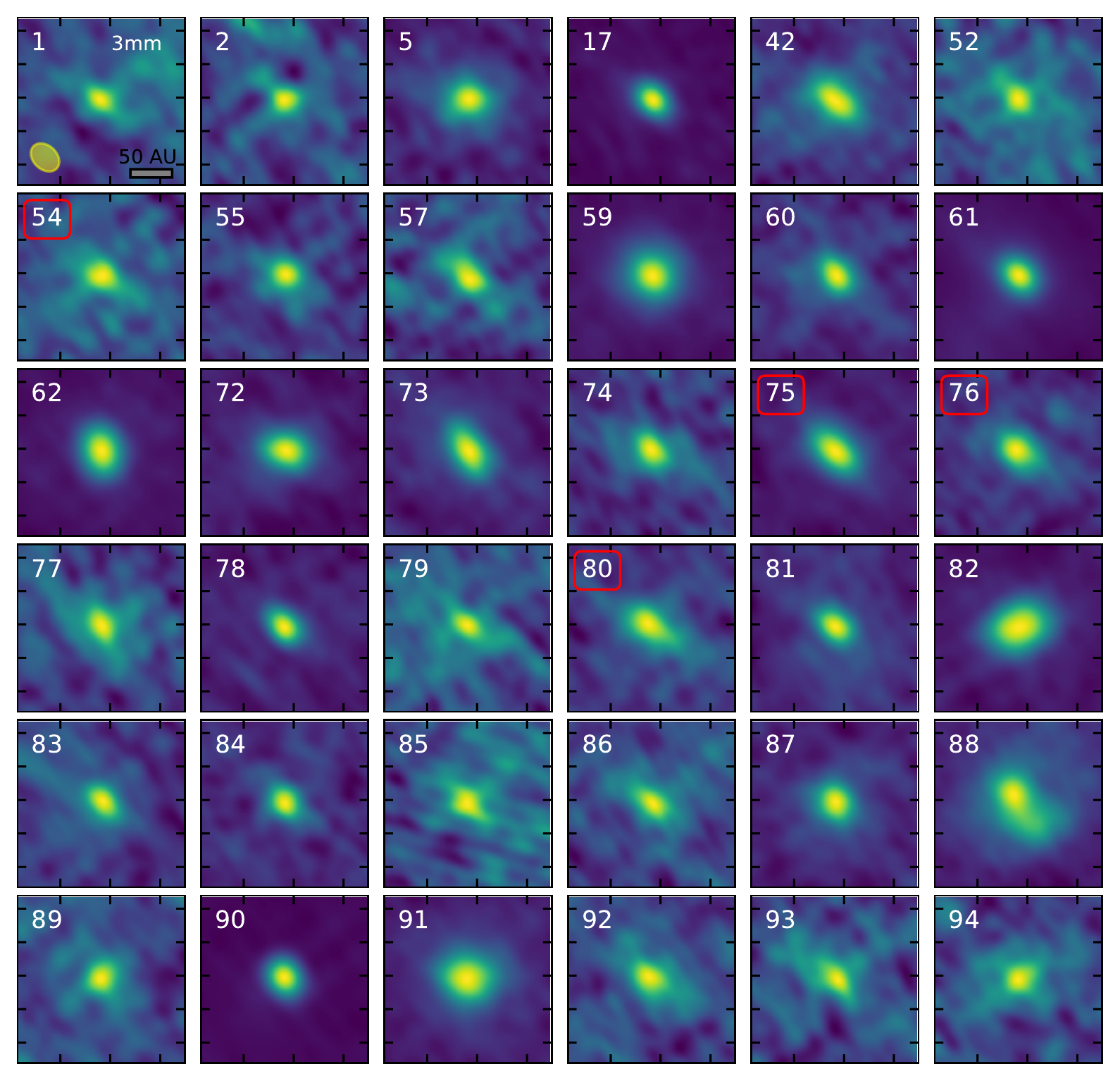}
    \caption{(1/2) Images of sources only in the band 3 FOV. The yellow ellipse shows the synthesized beam size. Each cutout is 0.5\arcsec\xspace across (200 AU). The number in the top left is the source ID, and source numbers boxed in red correspond to newly detected sources.} 
    \label{fig:b3_pg1}
\end{figure}
\setcounter{figure}{3}
\begin{figure}[!h] 
    \centering
    \includegraphics[width=0.7\textwidth]{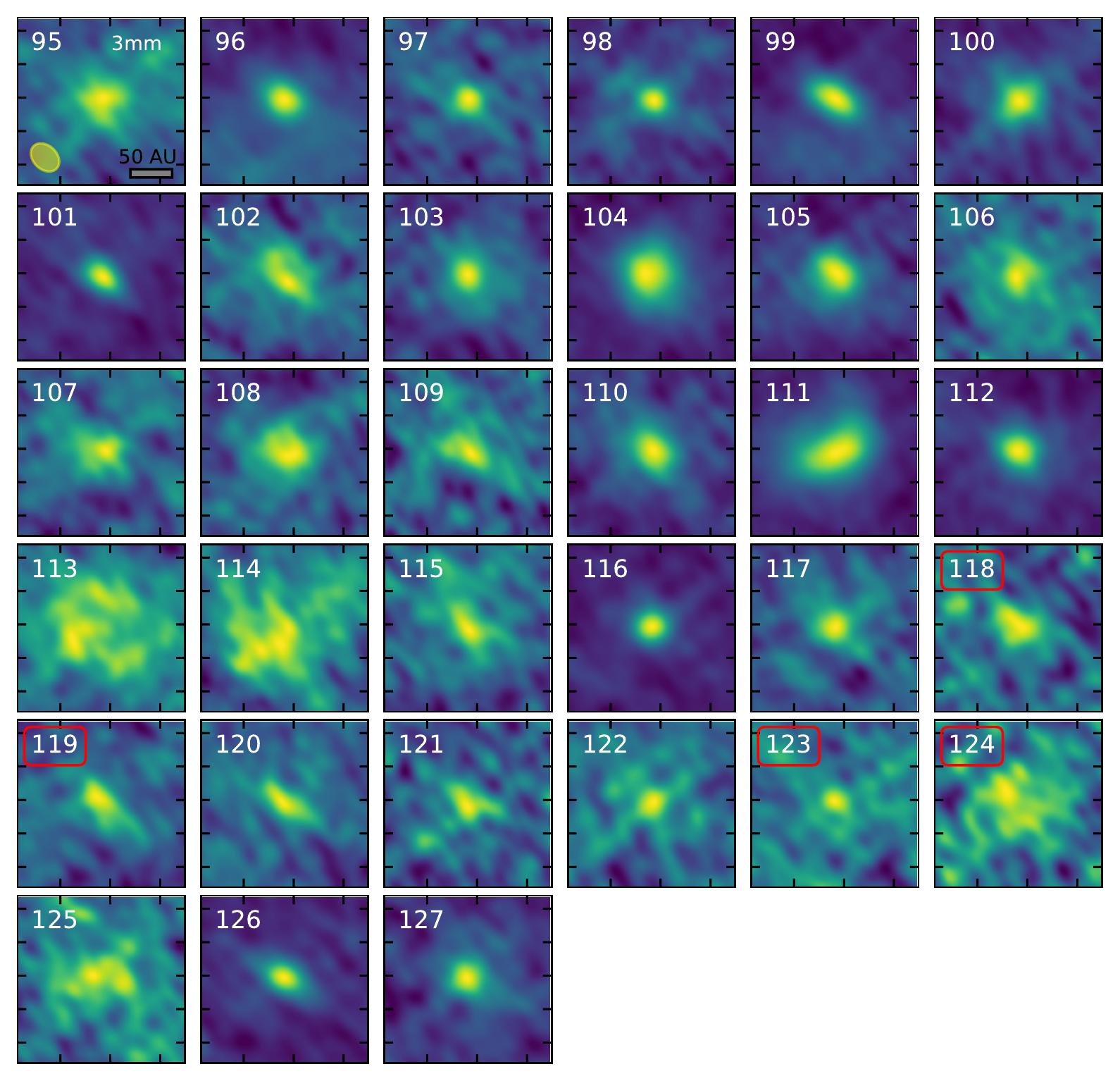}
    \caption{(2/2) Images of sources detected only in band 3. The yellow ellipse shows the synthesized beam size. Each cutout is 0.5\arcsec\xspace across (200 AU). The number in the top left is the source ID, and source numbers boxed in red correspond to newly detected sources.} 
    \label{fig:b3_pg2}
\end{figure}

\subsection{Correspondence with IR sources: the ONC and OMC1 samples} \label{sec:ir_dets}

We separate the sources into two groups: OMC1, those in or behind the molecular cloud, and ONC, those in front and associated with the nebula cluster.
We match our source locations to the near-IR (NIR) J, H, and K$_s$ band catalog of \citet{muench_luminosity_2002} (hereafter \citetalias{muench_luminosity_2002}), where we consider a NIR source a match if it is within 0.7\arcsec\xspace of the ALMA source, and find 76 matches, which are indicated in Table~\ref{tab:meas_misc_all}.
We compare our sample to more recent NIR studies with full coverage of our field of view such as \citet{robberto_wide-field_2010} and the VISION survey \citep{meingast_vision_2016} (both also J, H, and K$_s$ band studies), but these surveys have fewer matches (within the same 0.7\arcsec\xspace radius) than \citetalias{muench_luminosity_2002} with 60 and 62 matches respectively.
All the matches from \citet{robberto_wide-field_2010} and \citet{meingast_vision_2016} are included in \citetalias{muench_luminosity_2002} except for a match to source 70 in the VISION survey.

The resolution of the \citetalias{muench_luminosity_2002} data in the Trapezium region ranges from approximately 0.5\arcsec\xspace to 0.6\arcsec\xspace for different wavelength observations.
We choose 0.7\arcsec\xspace as the matching radius as slightly above the \citetalias{muench_luminosity_2002} resolution to account for the resolution of our data ($\sim$0.1\arcsec\xspace in band 3).
As the observations from \citetalias{muench_luminosity_2002} were conducted in March 2000 and our ALMA band 3 observations were taken in September 2017, we consider the potential impact of proper motions over this approximately 17 year epoch.
\citet{dzib_radio_2017} measure a mean proper motion of $\sim$1.3 mas yr$^{-1}$ for ONC stars, which yields a total movement of around 0.02\arcsec\xspace over this epoch.
BN is the source with the fastest proper motion of approximately 12 mas yr$^{-1}$ \citep[e.g.][]{dzib_radio_2017, goddi_multi-epoch_2011, rodriguez_proper_2005}, amounting to a distance of 0.2\arcsec\xspace over this time period, indicating that proper motions are significant for the fastest-moving sources.
Our results are not sensitive to the choice of matching radius: a radius as small as 0.3\arcsec\xspace yields 71 source matches, and a radius as large as 1\arcsec\xspace only adds 2 more source matches with 78 total matches.

Due to the high extinction of OMC1, we can generally distinguish sources in the ONC and OMC1 on whether they are IR detected.
Src I appears in the \citetalias{muench_luminosity_2002} catalog only for completeness and is not detected, so we include it in the OMC1 sample.
While BN may also be embedded in OMC1 \citep{bally_alma_2017} it is still IR detected despite high extinction so we include it in the ONC sample for consistency. 

There are several sources whose OMC1 classification is questionable.
We consider sources outside the dusty region shown in Figure~\ref{fig:gemini} where the majority of the OMC1 sources lie.
Sources 40 and 45 are both in a bright IR saturated region near BN and therefore may not be detected even if they are ONC sources.
Similarly, source 124 may be hidden by nearby bright IR sources, and is a spatial outlier in OMC1 (see the bottom left corner of Figure~\ref{fig:gemini}).
Source 22 is highly inclined, with a band 3 inclination of 85$^\circ$, and thus the disk may be thick enough to block the IR emission from the central star.

Within our band 3 field of view, we report 230 non-detections of IR sources from the \citetalias{muench_luminosity_2002} catalog in Table~\ref{tab:nondet_IR}.
We derive 3 mm flux upper limits for non-detected sources as three times the background RMS at each source location.
We compute the RMS following the method in Section~\ref{sec:analysis}, in an annulus of radius 0.1\arcsec\xspace with a width of 15 pixels.
We include these upper limits in Table~\ref{tab:nondet_IR}.

\begin{deluxetable*}{lll} \label{tab:nondet_IR}
\tablecaption{Nondetected IR sources. The IR identification is from the \citetalias{muench_luminosity_2002} catalog and alternate name from \citet{hillenbrand_constraints_2000}, with 230 sources in total.}
\tabletypesize{\scriptsize}
\startdata
& & \\
MLLA ID & Alternate ID & 3mm Flux Upper Limit (mJy)\\
386 & MS-66 & 0.39 \\
615 & IRc2 & 0.039 \\
336 & MS-1 & 3.9 \\
340 & MS-32 & 3.7 \\
346 &  & 1.4 \\
350 & MS-7 & 0.78 \\
\enddata
\tablecomments{Table~\ref{tab:nondet_IR} is published in its entirety in the machine-readable format. A portion is shown here for guidance regarding its form and content.}
\end{deluxetable*}

\subsection{Correspondence with other wavelength data} \label{sec:multiwave}

To better characterize the properties of these sources, we cross-match our catalog with the Chandra Orion Ultradeep Project (COUP) \citep{getman_chandra_2005}, a deep, sub-arcsecond resolution X-ray survey covering our field of view.
We find 70 source matches within 0.7\arcsec\xspace, 14 of which do not have corresponding IR detections.
YSOs emit X-rays primarily from hot plasma heated by magnetic flares from a central pre-main sequence star, so this relatively high detection rate (70/127 overall, 14/51 OMC1 sources) confirms there is a stellar component to many of our sources.
X-ray emission from YSOs is highly variable, so the lack of X-ray detections for the majority of these sources does not necessarily imply a lack of stellar activity, as the variability could drive the X-ray brightness below the sensitivity of \citep{getman_chandra_2005}.

We also compare our source detections with \citet{forbrich_population_2016}, a deep, high resolution (0.30\arcsec\xspace) centimeter survey of the ONC with the Karl G. Jansky Very Large Array (VLA).
We find 56 source matches within 0.7\arcsec\xspace, with 12 of these lacking IR detections.
The primary emission mechanisms for YSOs at centimeter wavelengths are thermal free-free emission and non thermal radio emission from stellar coronal activity, which also produces X-ray emission.
\citet{forbrich_population_2016} differentiates between these emission mechanisms by measuring spectral indices, where a negative spectral index indicates that the emission is primarily non-thermal.
5/12 of the matched, non-IR detected cm sources have measured spectral indices, and 1 of these 5 has a negative spectral index consistent with zero (source 49).
Of the four other cm detected sources with non-negative spectral indices, two are also x-ray detected (27 and 38), indicating that there is still stellar activity, though they have positive spectral indices. 

In total, 22/51 of our OMC1 non-IR detected sources have X-ray or cm detections.
While it is unclear whether the 8 radio detected sources that were not X-ray detected have coronal stellar activity, we still see that a significant fraction of our OMC1 sources have confirmed stellar activity. 
We compare the size and flux distributions of all ONC/OMC1 sources to those that are X-ray or cm detected and do not find any significant differences.

We again find that the choice of matching radius does not significantly impact our results.
For the cross-match to the COUP survey, changing our matching radius from 0.5 - 1\arcsec\xspace results in 70 - 71 matches.
Though this survey has slightly worse resolution than \citetalias{muench_luminosity_2002}, increasing the radius to 1\arcsec\xspace to account for this only adds one more source match, so we use the same matching radius for consistency.
Similarly, for the source matches to \citet{forbrich_population_2016}, we find the same change in matching radius from 0.5\arcsec\xspace to 1 \arcsec\xspace results in an increase in matches from 53 to 56 sources.
While we could decrease the matching radius to 0.5\arcsec\xspace for \citet{forbrich_population_2016} to reflect the higher resolution of these data, this would only remove 3 sources, so we use the same matching radius for simplicity.

The observations for the COUP survey and \citet{forbrich_population_2016} were performed in January 2003 and October 2012 respectively, corresponding to total sky movements of 0.18\arcsec\xspace and 0.06\arcsec\xspace for a source moving at a BN-like proper motion of 12 mas/yr.

\subsection{Newly Detected Sources} \label{sec:new_dets}

We find that 15 sources have not previously been reported by a SIMBAD and Vizier coordinate search, and were not included in any of the previous catalogs \citep[i.e.,][]{eisner_protoplanetary_2018, muench_luminosity_2002, getman_chandra_2005, forbrich_population_2016}.
We require that these sources are not matched in Vizier or SIMBAD to previously reported compact sources within a more stringent 1\arcsec\xspace radius.
To validate that these are genuinely new detections, we also consider whether nearby sources within 2\arcsec\xspace could be potential matches in Appendix~\ref{app:newdets}.

We detect two new candidate binary systems: sources 33 and 64, separated by $0.21 \pm 0.01$ arcseconds (84 $\pm$ 4 AU), and sources 118 and 119, separated by 0.58 $\pm$ 0.01 arcseconds (232 $\pm$ 4 AU).
We also resolve the IR binary source MLLA 689a/b (sources 54 and 56), which has angular separation 0.42 $\pm$ 0.2 arcseconds (168 $\pm$ 8 AU). 

\begin{figure}
    \centering
    \includegraphics{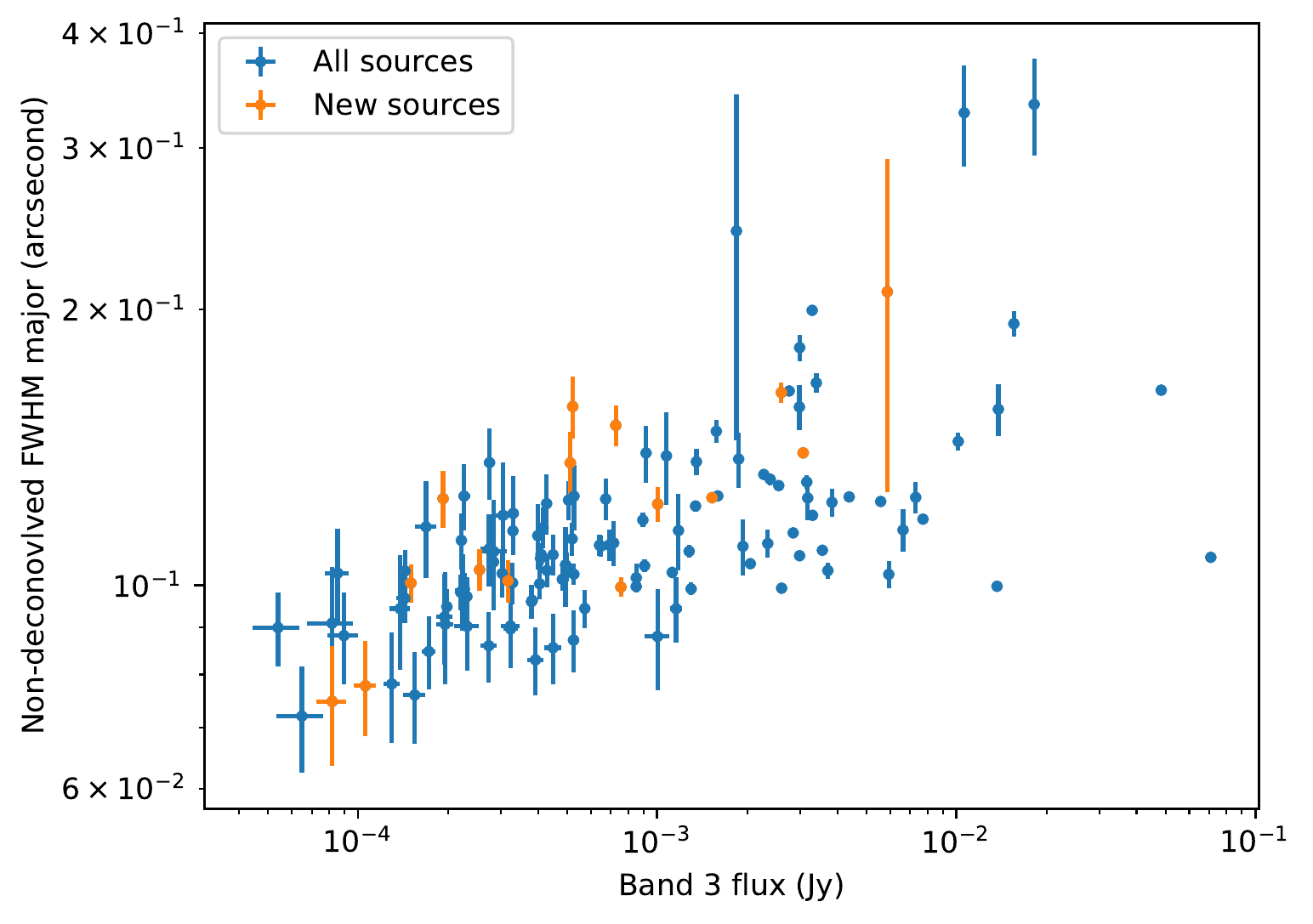}
    \caption{Band 3 non-deconvolved source FWHM and band 3 flux. Sources that are newly detected are plotted in orange, and other sources are in blue. The y-errors show the Gaussian fit uncertainties.}
    \label{fig:new_srcs}
\end{figure}

The newly detected sources are indicated in Table~\ref{tab:meas_misc_all}, and highlighted in Figures~\ref{fig:b3_pg1}, \ref{fig:b6_pg1}, and \ref{fig:b7_pg1}.
Because the target region is extremely well-studied at many wavelengths, including those studied here, it is surprising that we have detected so many new sources.
We do not find that the newly detected sources are particularly small or faint.
Figure~\ref{fig:new_srcs} shows the fitted (non-deconvolved) band 3 size and band 3 flux of these sources in orange and the rest of our sample in blue.
The newly detected sources span a range of sizes and fluxes, from 0.07\arcsec\xspace - 0.2\arcsec\xspace and 0.08 - 6 mJy, thus indicating that these new detections are not solely the result of greater sensitivity.
Five sources have band 3 fluxes of at least $>1$ mJy (32, 50, 75, 76, and 124) and should have been detected in previous surveys \citep[e.g.][]{eisner_protoplanetary_2016}, but were likely missed because of confusion with the hot core and other structures our observations have resolved out.

\subsection{Disk Fitting} \label{sec:fitting}

To obtain positions, sizes, and fluxes, we fit a 2D Gaussian to each source, then deconvolve the image beam from the Gaussian fit.
Source sizes are given by the major axis full width at half maximum (FWHM).
While we found that a 2D Gaussian fit the flux profile of most of our band 3 sources, the majority of band 6 and 7 sources are resolved and thus not well described with a Gaussian fit.
To enable fair comparison among the sizes of our sources in our different images, we convolve the band 6 and 7 images to the larger elliptical synthesized band 3 beam, effectively reducing the resolution.
We report source sizes measured from fits to the convolved images, then deconvolved by the band 3 beam.
Appendix~\ref{app:gauss} demonstrates why we adopt this process, showing an example of a source that has a greatly improved fit after the image convolution.
Eight band 6 sources and ten band 7 sources (5 in both bands) were undetectable when convolved to band 3 resolution; for these, we report fitted values derived from the original full-resolution images (these are indicated in our final catalog with $^\dagger$ for band 6 and $^*$ for band 7). 

Three band 3 sources (52, 71, and 77) have background extended emission that interferes with the source fitting, so we fit a second, broader Gaussian component simultaneously as a background model.
We also found that 4 band 3 sources (113, 114, 124, and 125) have very poor Gaussian fits, but were large enough to be deconvolved, so we do not include these deconvolved sizes in our analysis, though we report the sizes in our catalog.

We determine whether the sources were resolved or not by comparing their fitted Gaussian size with the synthesized beam.
Resolved sources are ``deconvolvable" if the fitted Gaussian is larger than the synthesized beam.
More detail on this criteria is provided in Appendix~\ref{app:size}.
For sources detected in the band 6 and 7 data convolved to band 3 resolution, we compare their  sizes after convolution to the band 3 beam, and for sources detected only in the original band 6 and 7 images, we deconvolve the original band 6 or 7 beam.
We find 17/34 sources are resolved in band 7, 31/50 sources in band 6, and 72/127 sources in band 3.
13 sources were resolved in all three bands.

In addition to measuring fluxes from the Gaussian fits, we measure aperture fluxes by summing pixel values within an elliptical aperture with semi-major and minor radii of the fitted source FWHM major and minor, converting between Jy/beam and Jy/pixel by dividing by the number of pixels in the synthesized beam.
We then estimate the background by subtracting the median flux in a circular annulus surrounding the source. 
This annulus has an inner radius of the source major FWHM, and an outer radius of the source FWHM plus 15 pixels. 
We calculate the error of the flux measurement by calculating the RMS in the annulus.
Finally, we add an aperture flux correction of 1.85\%, appropriate for a Gaussian truncated at twice the FWHM. 
We note that flux calibration for ALMA has an additional uncertainty as high as $\gtrsim$10\% \citep{francis_accuracy_2020}.
Unless otherwise stated, we use the measurement uncertainties described above only.
The aperture fluxes agree with fluxes computed directly from the Gaussian fits within 10\% for the majority of sources.
The aperture flux is a more direct flux measurement for our resolved sources that do not have perfectly Gaussian flux profiles (such as Src I), so we use the aperture flux for all sources for consistency. 


\subsection{Disk Measurements}\label{sec:diskmeasurements}

\begin{figure}[!h]
    \centering
    \includegraphics[width=0.9\textwidth]{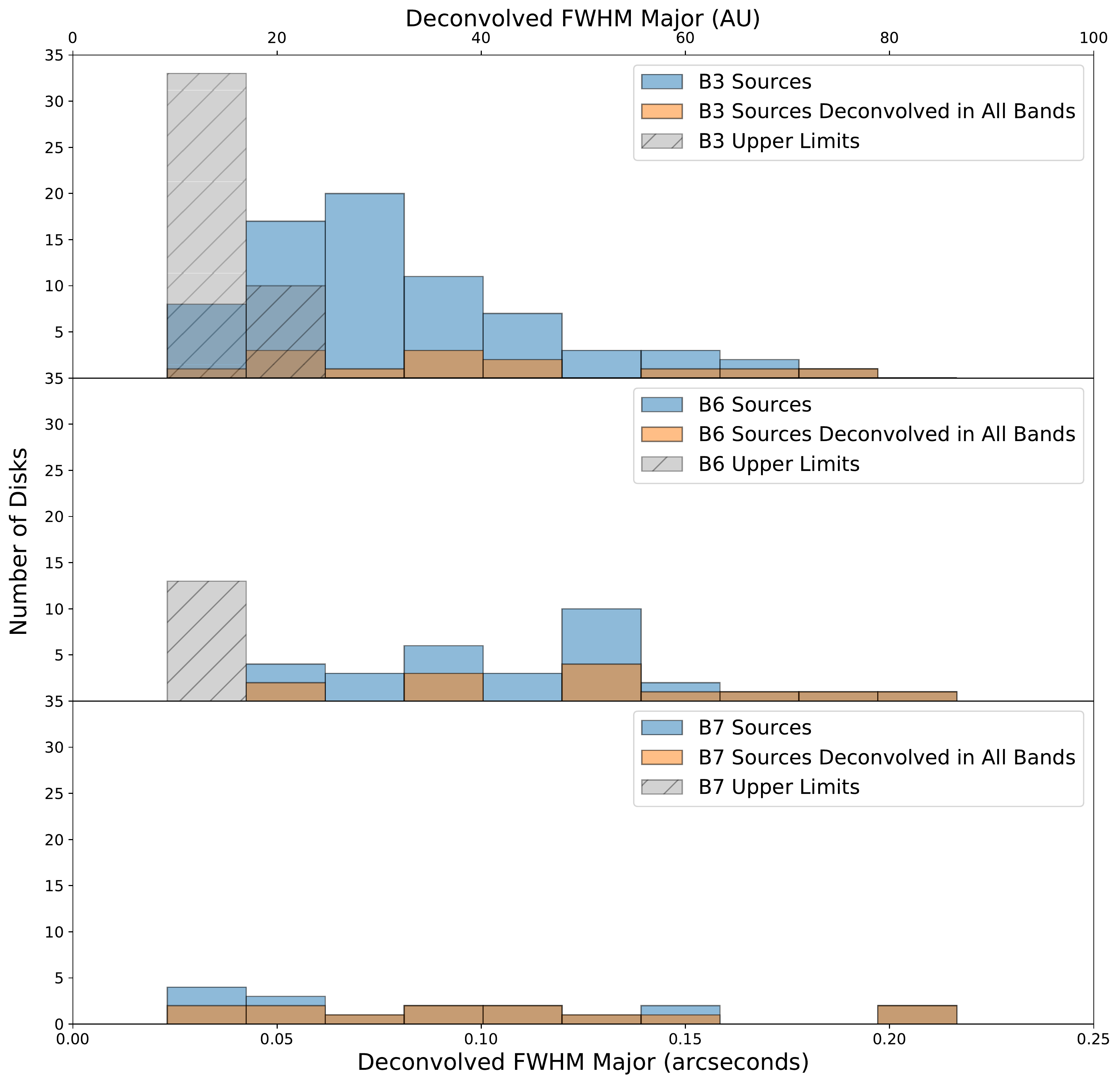}
    \caption{Histograms of the deconvolved major FWHM of sources at 3, 1.3, and 0.85 mm (bands 3, 6, and 7)} from top to bottom. The orange bars are sources that are resolved in all three bands, whereas blue bars are sources that are not. The hatched gray bars show upper limit sizes of unresolved sources.
    \label{sizehist}
\end{figure}

Figure~\ref{sizehist} shows a histogram of the deconvolved major FWHM of disks in bands 3, 6, and 7.
In band 3, the majority of deconvolvable sources are smaller than $\sim45$ AU.
The band 6 and 7 sizes span a similar size range but are less strongly peaked at small sizes.

To derive size upper limits on unresolved sources, we find the minimum radius at which a uniform disk can be deconvolved as a function of S/N for each image (median upper limit sizes are 14, 10, and 4 AU in bands 3, 6, and 7 respectively).
Appendix~\ref{app:size} describes this experiment in detail.

At (sub-)mm wavelengths, we expect in general to observe larger disk sizes at shorter wavelengths.
There are two effects that control the apparent disk size: noise and opacity.
For an optically thin disk, at larger radii, where the column density is lower, the disk will eventually blend into the noise background and become undetectable.
Because of the steep dust opacity index, this effect will make disks appear larger at shorter wavelengths if observed with the same sensitivity.
If the inner disk is optically thick at short wavelengths but optically thin at long wavelengths, it will tend to look more centrally peaked at long wavelengths.
Both of these effects tend to make disks appear larger at shorter wavelengths.

\begin{figure}[!h]
    \centering
    \includegraphics[width=0.45\textwidth]{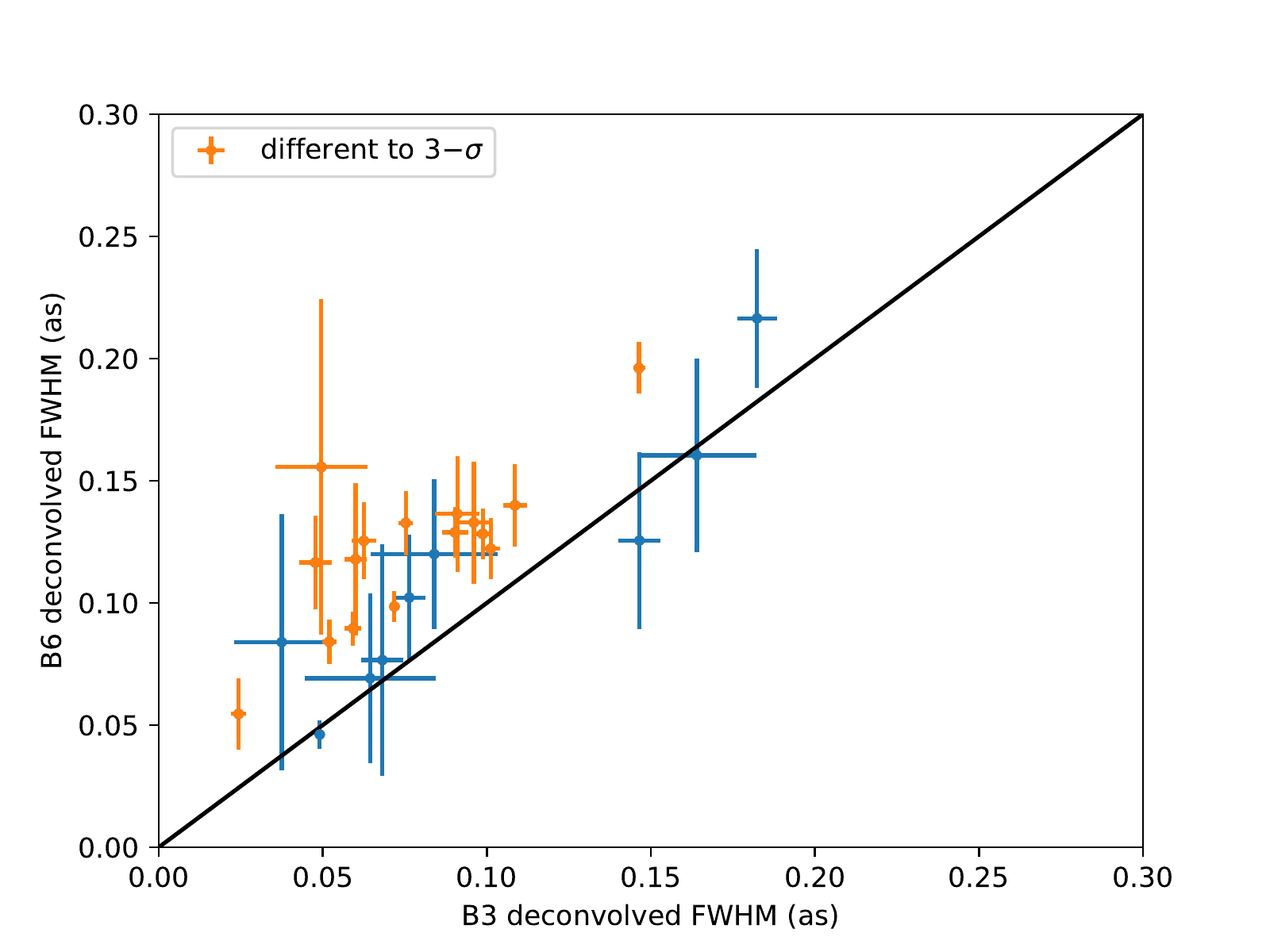}
    \includegraphics[width=0.45\textwidth]{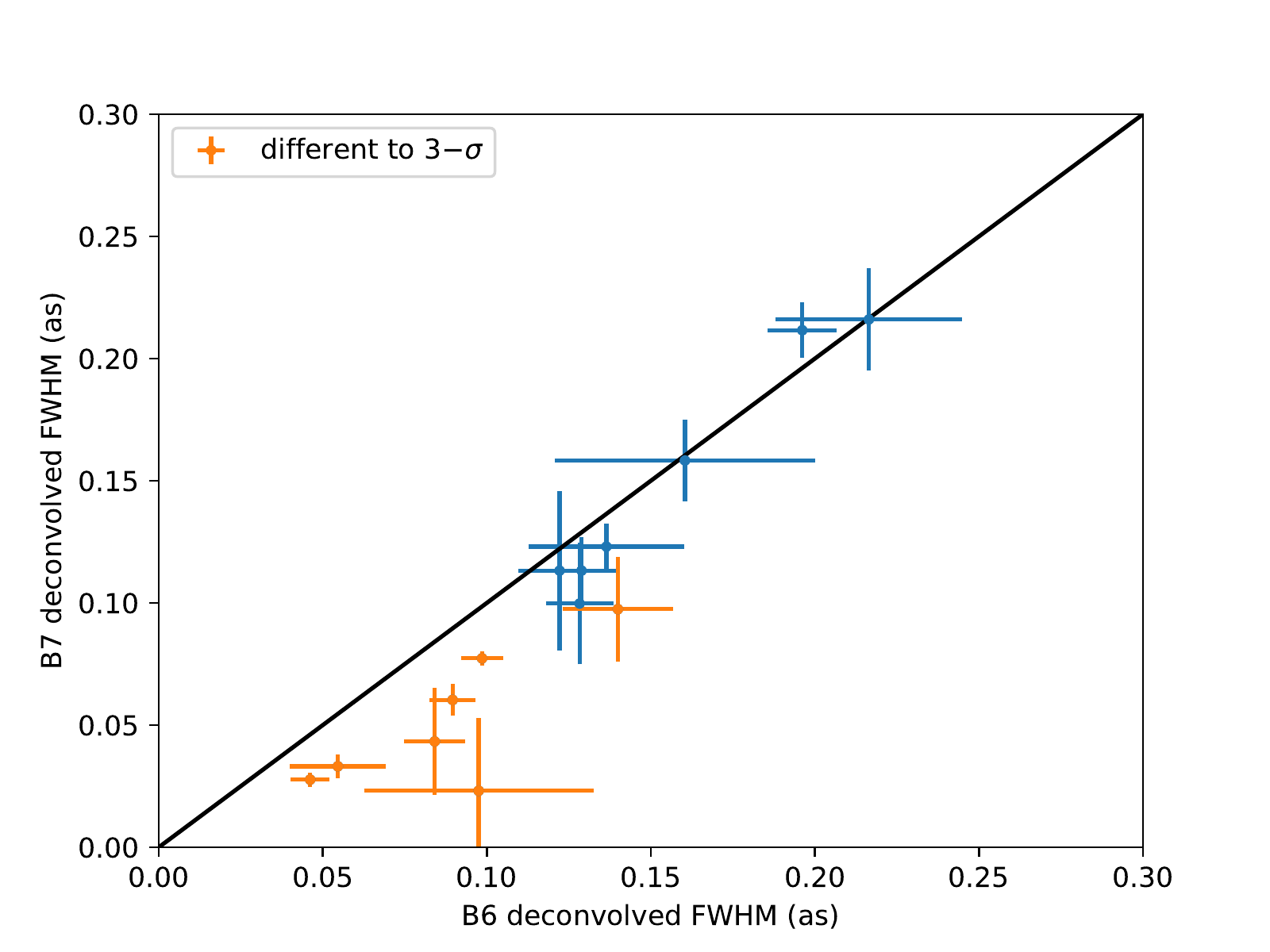}
    \caption{Plots comparing the sizes of sources resolved in multiple bands. Left: sources resolved in bands 3 and 6. The sizes are the deconvolved major FWHM with 3-$\sigma$ error bars. Sources that have significantly different sizes ($>3\sigma$) are shown in orange. The black line shows where the band 3 and band 6 sizes are equal. 25 sources are shown. Right: The same for bands 6 and 7. 14 sources are shown.}
    \label{band_size}
\end{figure}

We compare the relative sizes of sources at different wavelengths in Figure~\ref{band_size}.
The left panel shows the deconvolved major FWHM of sources in bands 3 and 6 (3 and 1.3 mm respectively), where we see that band 6 sizes are systematically larger. 
We find 16 sources are significantly (to 3$-\sigma$) larger in band 6 than band 3, and 5 of those 16 were significantly larger in band 7 than band 3. 
Additionally, only source BN (39) is apparently larger in band 3 than band 7.
On the right, we see that sources have similar band 6 and 7 sizes, though 7 sources are significantly larger in band 6 (27, 28, 29, 31, 32, 34, and BN). 

We infer the inclination, $i$, from the ratio of the minor axis to the major axis, assuming an infinitely thin disk (but we test this assumption below), hence $i = 0^\circ$ corresponds to a face-on disk and $i = 90^\circ$ means the disk is edge-on.
If the disk has finite width, the minimum minor/major ratio is the scale height of the disk divided by its effective radius, which systematically decreases the inferred inclination, affecting edge-on disks more than face-on.

\begin{figure}
    \centering
    \includegraphics[width=0.8\textwidth]{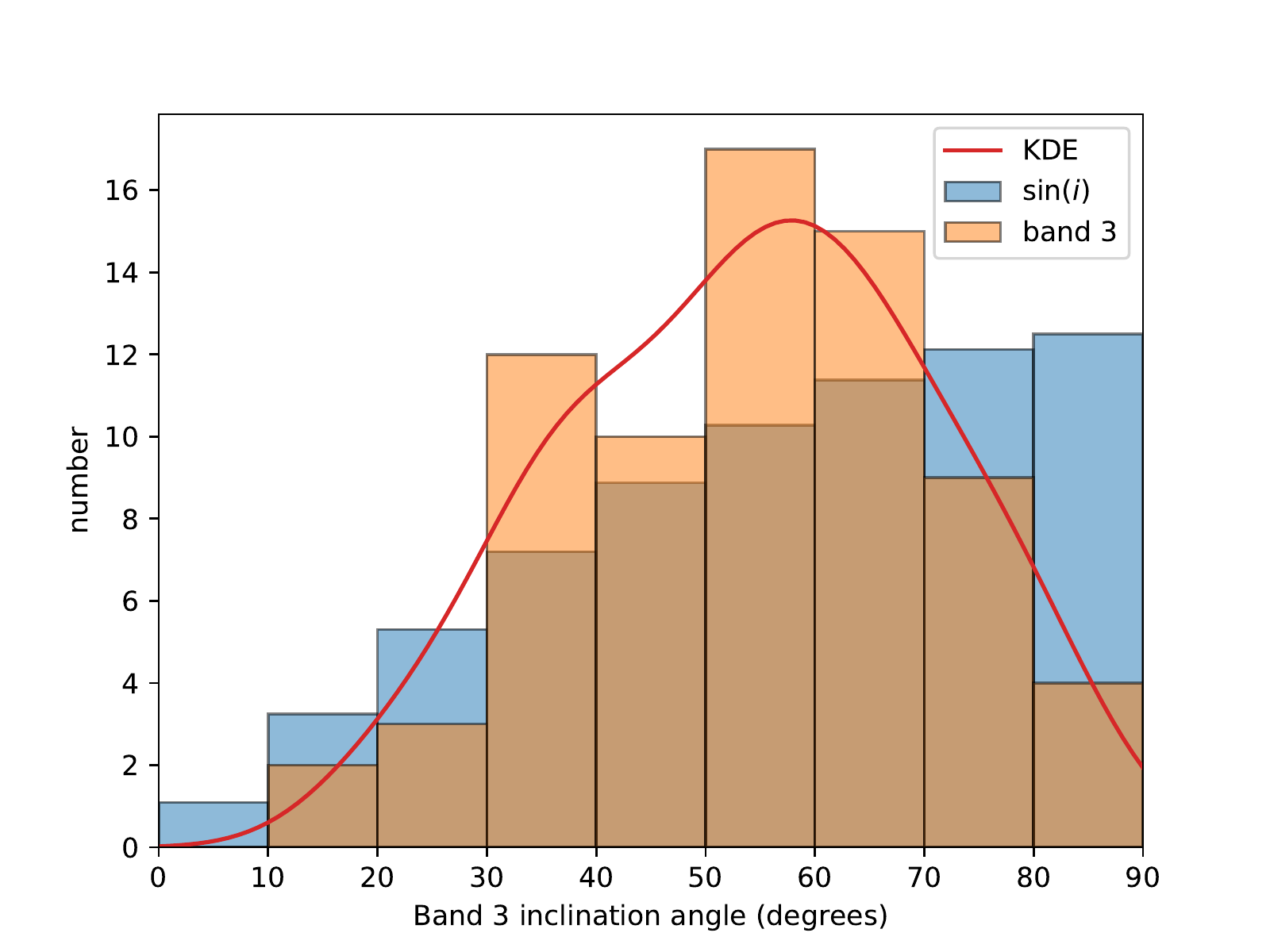}
    \caption{A histogram of band 3 inclination angles (orange bars). For the disks to appear at random alignments, we would expect the probability of each angle $p(i) \propto \sin(i)$ ($p(\cos i) \propto const$), which is shown in blue bars. A Gaussian KDE of our data with bandwidth 0.49 is shown as a red line, to guide the eye.}
    \label{incl_hist}
\end{figure}

In Figure~\ref{incl_hist}, we plot a histogram of the measured band 3 inclinations with a Gaussian Kernel Density Estimator (KDE) of our data in red with a bandwidth of 0.49 \citep[calculated according to Scott's rule,][]{scott_multivariate_1992}, drawn to guide the eye. 

We expect the probability of a disk inclination angle to be proportional to $\sin(i)$ for randomly aligned disks\footnote{If a vector has an equal likelihood of pointing in any direction, in a spherical coordinate system $\int P(\theta) d\phi \sin\theta d\theta = 1$ where $P(\theta)$ is the probability the vector points at some angle $\theta$. Then, $P(\theta) = \frac{1}{\pi}\cos\theta$, and $i = 90 - \theta$, so $P(i) \propto \sin i$.}.
We plot the expected histogram if the disks followed this distribution exactly in Figure~\ref{incl_hist}.
We perform a Kolmogorov-Smirnov (KS) test between the two distributions and find a p-value of 0.008, thus we reject the null hypothesis that the underlying distributions are the same.
We suggest the difference is due to finite vertical height of the disks; contrary to our assumption of a thin disk.
This nonzero width of the disk makes inclined sources appear less inclined, leading to fewer sources at high inclinations as we observe in Figure~\ref{incl_hist}.

We do not find an excess of high-inclination disks in our OMC1 sample, indicating that most of these YSOs are truly embedded rather than foreground sources obscured by a highly inclined disk.


\subsection{Spectral Indices} \label{sec:alpha}

\begin{figure}
    \centering
    \includegraphics[width=\textwidth]{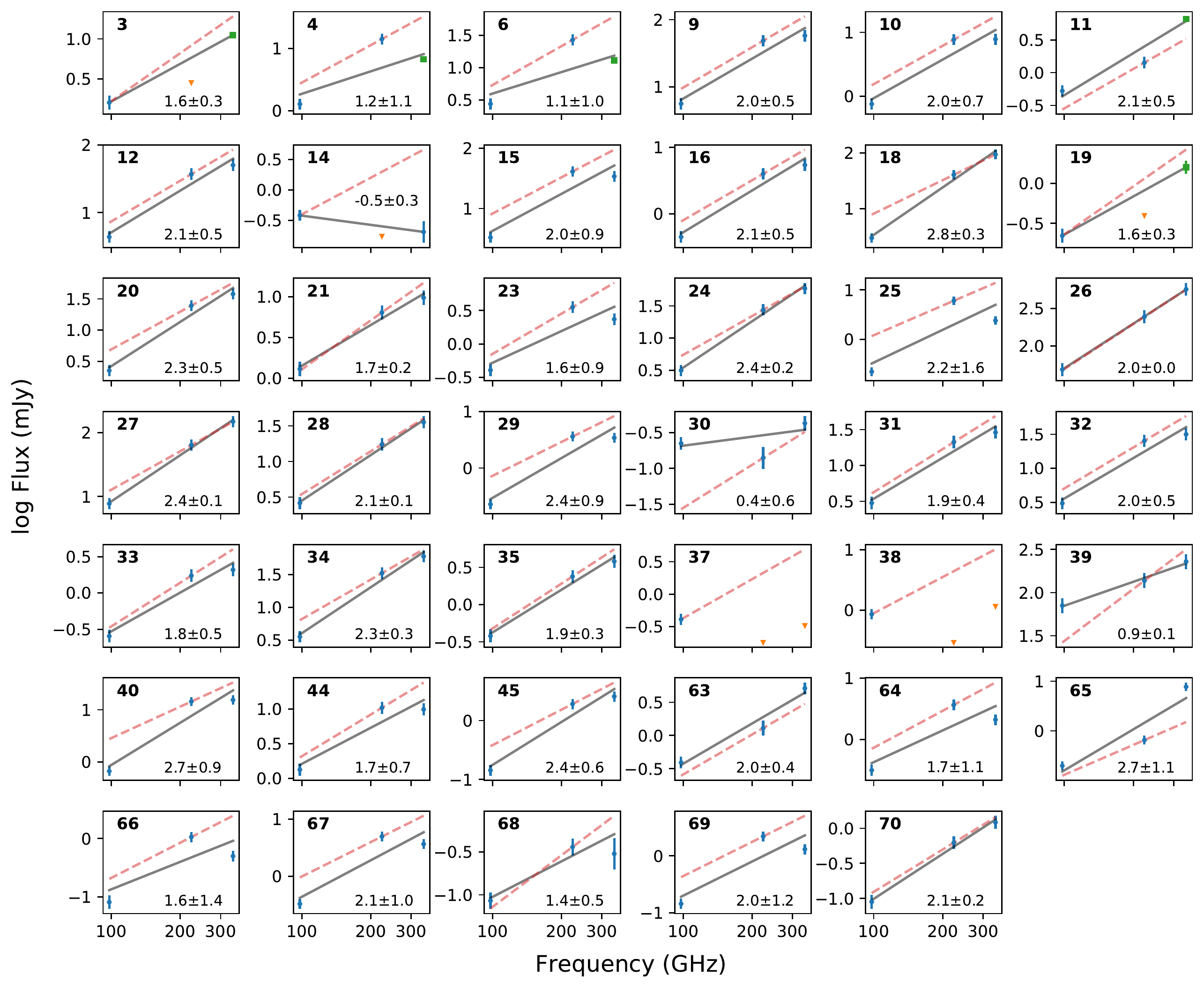}
    \caption{Our measured SED for every source in the band 7 field of view. Blue points are detections in our data, green squares are fluxes from \citetalias{eisner_protoplanetary_2018}, and orange triangles are flux upper limits. The number in the top left of each panel is the source number. The red dashed line shows an SED slope of 2, normalized to the band 6 flux, and the solid black line is a power-law fit, with the slope shown in the bottom right. Uncertainties include an additional 20\% flux calibration contribution (see Section~\ref{sec:fitting}).}
    \label{fig:seds}
\end{figure}

Figure~\ref{fig:seds} shows the SEDs of sources with detections or upper limits in all three bands.
The red dashed line overlaid shows an SED slope with spectral index $\alpha = 2$ in red normalized to the band 6 flux when available, or the band 3 flux otherwise.
The black line shows a power-law fit to the data, with the measured spectral index shown on the bottom of each panel.
We include a conservative 20\% flux calibration uncertainty in addition to measurement uncertainties for spectral index measurements, as mentioned in Section~\ref{sec:fitting}.
We include band 7 fluxes from \citetalias{eisner_protoplanetary_2018} for sources 4, 6, and 11 which have band 3 and 6 detections but are outside the field of view of our band 7 data.

Source 15 is the only source that is detected in both \citetalias{eisner_protoplanetary_2018} and our band 7 data.
We find that our measured band 7 flux ($34\pm 7$ mJy, including 20\% systematic uncertainty) agrees with that of \citetalias{eisner_protoplanetary_2018} ($17.3 \pm 2.0$ mJy) within 3--$\sigma$.

In the overlap region of \citetalias{eisner_protoplanetary_2018} and our band 3 data, we detect 39/56 sources.
35 of these detected sources are only in our band 3 field of view, so we utilise the flux measurements of \citetalias{eisner_protoplanetary_2018} to measure spectral indices for these sources.
Additionally, we measure band 3 flux upper limits (as in Section~\ref{sec:ir_dets}) for the \citetalias{eisner_protoplanetary_2018} sources we fail to detect, yielding spectral index lower limits for these sources.

In total, we measure the spectral index for 86 sources: 37 sources fit with three flux measurements, 14 sources with only band 3 and band 6 or 7 measurements, and 35 sources with band 3 and \citetalias{eisner_protoplanetary_2018} fluxes.
We measure upper limits on the spectral index for six band 3 sources that we fail to detect in our band 6 data, and lower limits for the 17 \citetalias{eisner_protoplanetary_2018} we do not detect in our band 3 data.
In Table~\ref{tab:meas_misc_all}, we report our measured spectral indices between band 3 and band 6 ($\alpha_{B3\to B6}$), between band 6 and band 7 ($\alpha_{B6\to B7}$), and our best estimate spectral index ($\alpha$): the fitted spectral index for sources shown with three flux measurements, or a two band spectral index measurement for sources with only two flux measurements.

\begin{figure}
    \centering
    \includegraphics[width=0.8\textwidth]{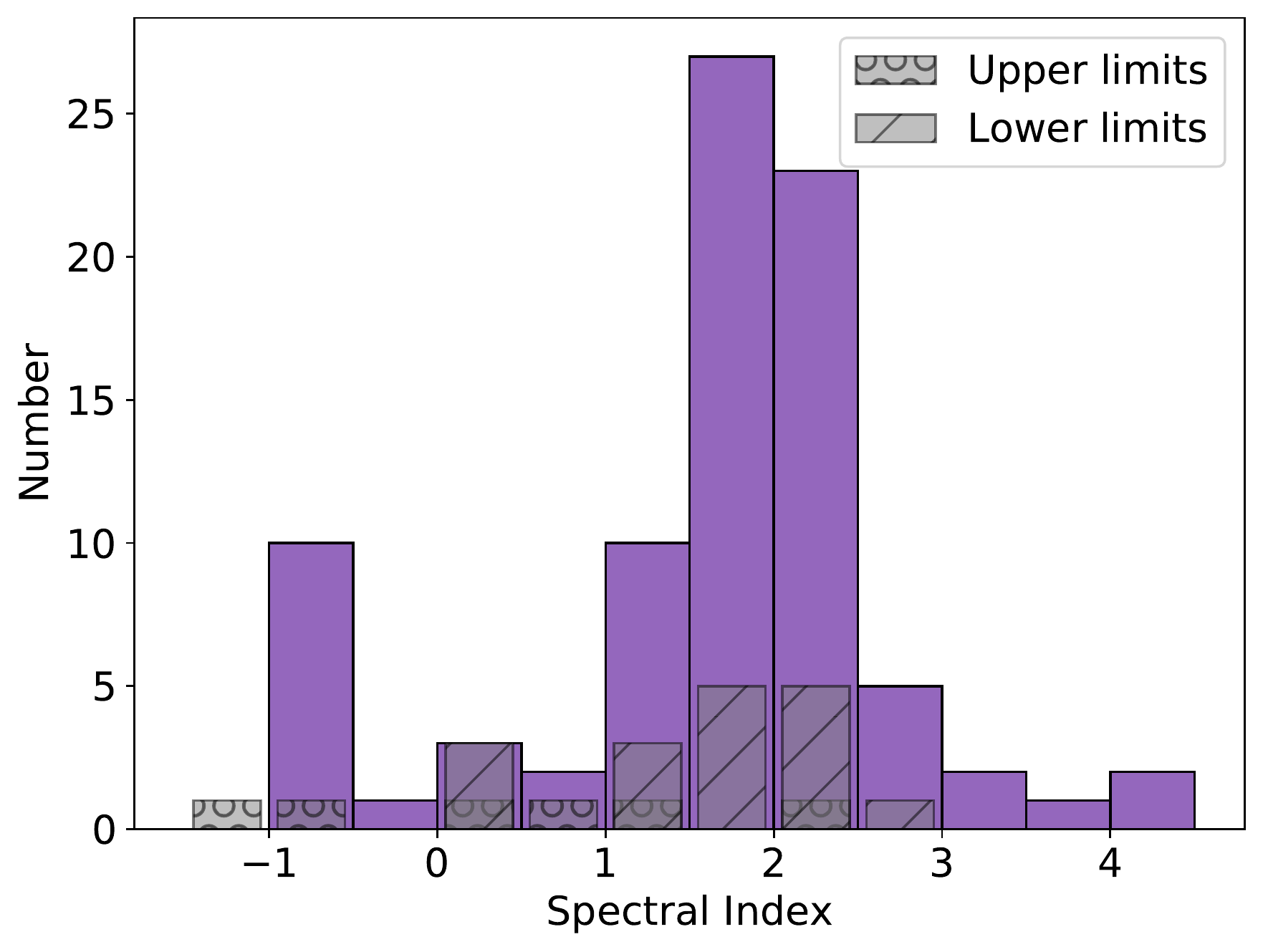}
    \caption{A histogram of spectral indices for the 85 sources with multiple flux measurements in purple. 37 sources have fitted spectral indices as shown in Figure~\ref{fig:seds}, and the remaining 49 sources have only two flux measurements. Gray bars with the circle pattern correspond to upper limit spectral indices (band 3 detection and band 6 non-detection), and gray diagonally hatched bars show lower limit spectral indices (band 3 non-detection and \citetalias{eisner_protoplanetary_2018} detection).}
    \label{fig:alpha_hist}
\end{figure}

Figure~\ref{fig:alpha_hist} shows a histogram of all sources with measured spectral indices (the best estimate described above) and upper and lower limits.
The distribution of spectral indices has a clear peak around $\alpha=2$, a tail towards higher spectral indices, and another peak at $\alpha=-1$.
We find that 59/86 sources have spectral indices consistent with $\alpha=2$ within 3--$\sigma$ uncertainties (i.e. $\alpha=2$ is contained in the 3--$\sigma$ interval), and of the remaining 27 sources, 5 have $\alpha > 2$ and 22 have $\alpha < 2$.

As discussed in \citet{ballering_protoplanetary_2019}, the disk spectral indices vary with parameters such as disk size and dust mass, so it is not straightforward to infer the optical depth based on the spectral index.
In their models, if the disk is entirely thin then $\alpha \gtrsim 2.7$, and models with partially optically thick disks have $\alpha>2.4$.
In total, 50/86 sources have $\alpha$ consistent with $\alpha\geq 2.4$ within 3--$\sigma$ uncertainties, but the most common measurement is close to $\alpha\approx2$, hinting that the disks may be very optically thick.

There are 3 disks with linearly fitted SEDs (i.e. with 3 flux measurements) that have $\alpha < 2.4$ to 3--$\sigma$ (i.e. the entire 3--$\sigma$ interval is below 2.4): 21, Src I (26), and BN (39).
Source 21 and Src I have $\alpha \approx 2$, while BN has a much flatter $\alpha = 0.9 \pm 0.1$.

Sources with $\alpha < 0$ may have a significant flux contribution from free-free emission.
\citetalias{eisner_protoplanetary_2018} report disk fluxes after modelling and removing free-free emission contributions (though we use the total disk fluxes in the analysis above).
13/34 \citetalias{eisner_protoplanetary_2018} sources included here have free-free contributions greater than the 1--$\sigma$ flux uncertainty, and 7 are determined to be entirely dominated by free-free emission.
All of these 13 sources have $\alpha < 2$ to 3--$\sigma$, and 9 have $\alpha < 0$ to 3--$\sigma$.
No other sources have $\alpha < 0$ to 3--$\sigma$.

\subsection{Dust masses} \label{sec:dmasses}

To estimate dust masses of the disks, we use Equation~1 from \cite{eisner_protoplanetary_2018}:
\begin{equation} \label{eq:dmass}
    M_{dust} = \frac{S_{\nu, dust} d^2}{\kappa_{\nu, dust} B_\nu(T_{dust})},
\end{equation}
where $S_{\nu, dust}$ is the flux from dust, ${\kappa_{\nu, dust} = \kappa_0\left(\frac{\nu}{\nu_0}\right)^\beta}$ is the dust mass opacity, $d$ is the distance, $T_{dust}$ is the dust temperature, and $B_\nu$ is the Planck function.
We assume optically thin dust, $d = 400$ pc, $\kappa_0 = 2$ cm\textsuperscript{2}/g at 1.3 mm ($\nu_0 = 230$ GHz), $\beta = 1.0$, \citep{hildebrand_determination_1983, beckwith_particle_1991} and $T_{dust} = 20$ K \citep[e.g.][]{carpenter_2mass_2000, andrews_circumstellar_2005}.
These dust opacity and temperature values are the same as those of \citetalias{eisner_protoplanetary_2018}.
We find a range of masses from $\sim$ 1 M$_\oplus$ to $\sim$ 4000 M$_\oplus$ (Src I). 
Because we have assumed the dust is optically thin, our mass estimates are lower limits.
However, at least 13 sources (as discussed in Section~\ref{sec:alpha}) have significant flux contributions from free-free emission, so their dust masses are instead overestimated with Equation~\ref{eq:dmass}.

\section{Distinguishing the OMC1 and ONC populations} \label{sec:oncvsomc1}

\subsection{ONC and OMC1 Disk Properties}

In this section, we compare disk properties in our ONC and OMC1 samples and find similar disk sizes but differing flux distributions.
To expand our ONC sample, we incorporate the sources in \citetalias{eisner_protoplanetary_2018}.
Because \citetalias{eisner_protoplanetary_2018} only search for sources within 0.5\arcsec\xspace of optical or near-IR source locations, their sample is selected similarly to our IR-detected ONC sample, with a sensitivity comparable to our band 7 data ($\sim$0.1 mJy).

\begin{figure}
    \centering
    \includegraphics[width=0.9\textwidth]{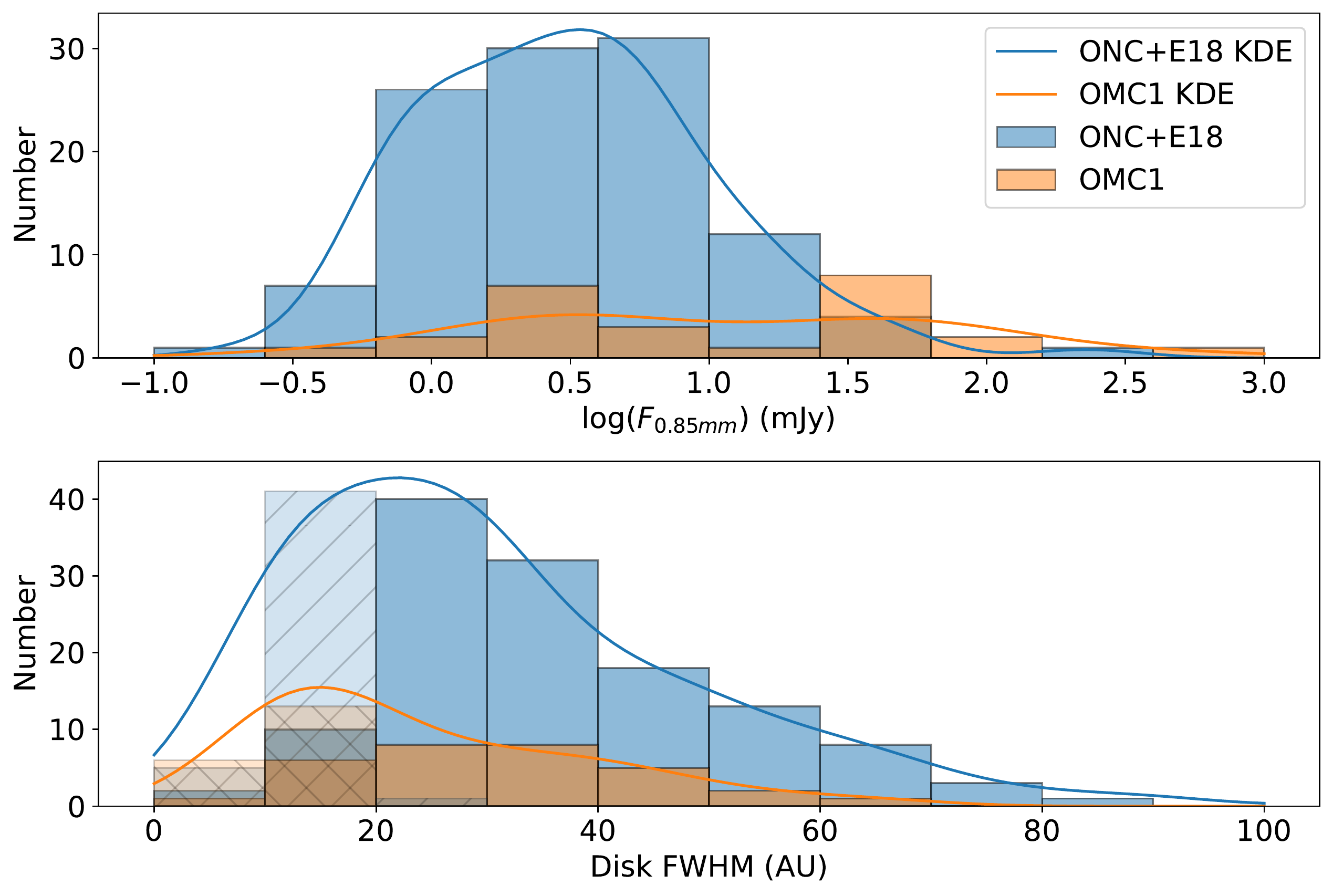}
    \caption{Above: A band 7 (0.85 mm) flux histogram of the combined ONC sample (including \citetalias{eisner_protoplanetary_2018}) in blue bars, and the OMC1 fluxes in orange bars.
    Gaussian KDEs are drawn to guide the eye, with bandwidths of 0.39 and 0.53 for the ONC (blue) and OMC1 (orange) samples respectively.
    Bottom: a size histogram for the combined ONC and OMC1 samples with the same coloring as above. Transparent hatched bars correspond to upper limit sizes for unresolved sources. Sizes for our sources are measured in the band 3 data (3 mm). Associated Gaussian KDEs have bandwidths of 0.36 and 0.46 for the ONC and OMC1 samples, and include the upper limit sizes.}
    \label{fig:onc_omc1_hist}
\end{figure}

The flux distribution of the combined ONC sample (including \citetalias{eisner_protoplanetary_2018}) is different from that of the OMC1 sample, while their size distributions are similar.
The upper panel of Figure~\ref{fig:onc_omc1_hist} shows a band 7 flux histogram of the combined ONC sample and our OMC1 sample, with Gaussian KDEs drawn to guide the eye (bandwidths of 0.39 and 0.53 respectively, calculated with Scott's rule as above).
We see that the OMC1 sample has an excess of bright sources and a relatively uniform distribution, compared to the combined ONC distribution which peaks at around $\sim$3 mJy and contains only one source above 100 mJy, source BN.
To test whether the qualitatively different shapes of the flux distributions is statistically significant, we use the KS-test, which yields a significant p-value of 0.002.
The excess of bright sources and large flux range in our OMC1 sample indicates that it is physically different from the ONC sample.

\begin{figure}
    \centering
    \includegraphics[width=0.6\textwidth]{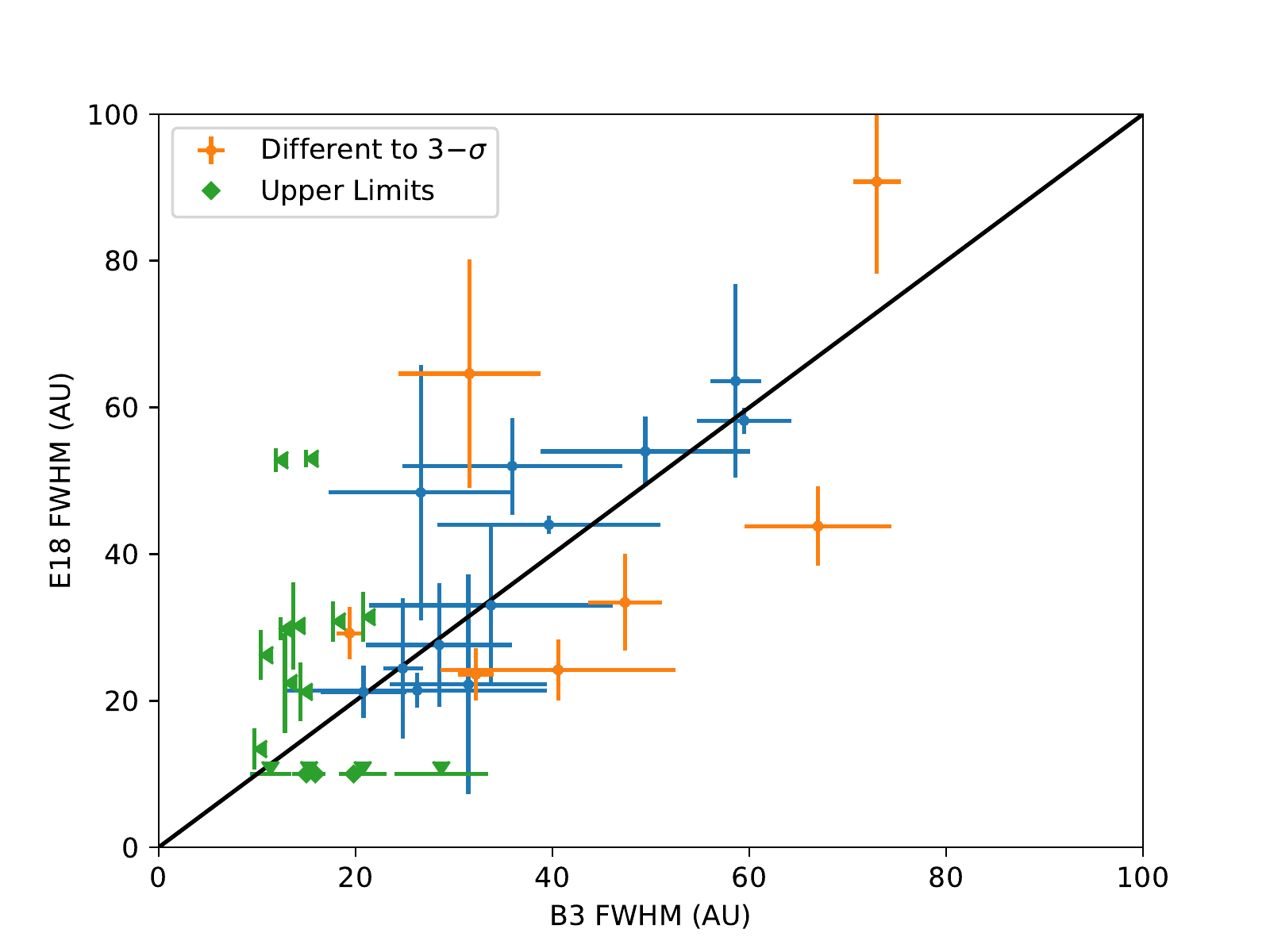}
    \caption{Measured band 3 and \citetalias{eisner_protoplanetary_2018} (band 7) sizes for sources detected in both studies. Points in blue are consistent within 3-$\sigma$, while orange points have different sizes. Green triangles show sources with size upper limits in one or both studies.}
    \label{fig:eis_size}
\end{figure}

The source size distributions of the combined ONC and the OMC1 samples are not distinguishable.
In the lower panel of Figure~\ref{fig:onc_omc1_hist}, we plot the size distributions, with KDEs of bandwidths of 0.36 and 0.46 respectively.
Because only 17 sources are resolved in our band 7 data, we plot the band 3 sizes.
While in our own data we find that band 7 sizes are typically slightly larger for sources detected in both band 3 and 7, when comparing the sizes of \citetalias{eisner_protoplanetary_2018} sources detected in our band 3 data, we do not see a systematic offset.
Figure~\ref{fig:eis_size} shows the band 3 and \citetalias{eisner_protoplanetary_2018} band 7 sizes for the 39 sources detected in both studies, of which 19 are resolved in both studies.
Though 7 sources have significantly different (to 3-$\sigma$) sizes, there is no clear offset.
Then, we find that the comparison of our band 3 and \citetalias{eisner_protoplanetary_2018}'s band 7 sizes in Figure~\ref{fig:onc_omc1_hist} is not likely to be a significant source of uncertainty.
Indeed, we find that the size distributions for the combined ONC and OMC1 samples are similar, and cannot be distinguished with a KS-test (p-value of 0.37).

\bigskip \bigskip \bigskip \bigskip
\subsection{Source Density}

We find evidence that the optically obscured OMC1 sample comprises a spatially distinct subcluster from the broader ONC population.

\begin{figure}
    \centering
    \includegraphics[width=\textwidth]{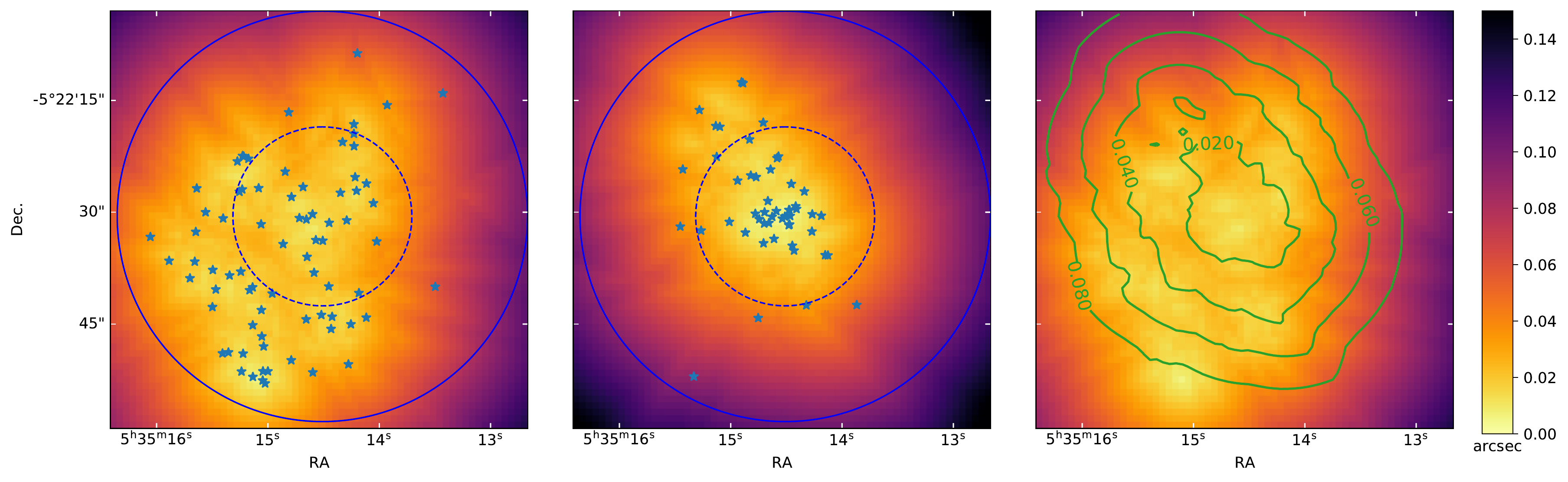}
    \caption{Left: A map of our band 3 FOV with our ONC (IR detected) source locations shown with blue stars. The colormap (in units of arcseconds) shows the 5th nearest neighbor to each pixel position. The dashed blue circle is the FWHM of the band 3 primary beam, and the solid blue circle is the band 3 field of view (5\% PB recovery). Center: The same but for our OMC1 sample (not IR detected). Right: The colormap shows the 5th nearest neighbor distance for the ONC sample (same as left panel), and the contours show the same for the OMC1 sample. The OMC1 sample is more concentrated.}
    \label{fig:density_3panel}
\end{figure}

We analyze the clustering of the disk-bearing samples in the ONC and OMC1.
Figure~\ref{fig:density_3panel} shows the source locations and densities of the ONC (left) and OMC1 (center) samples.
The density is estimated by determining the distance from each pixel to the 5th nearest neighbor star in a 64 by 64 grid spanning the band 3 field of view.
Source locations are shown with blue stars, and the colorbar is in units of arcseconds. 
While the results are similar for different $n$th nearest neighbors, we choose to calculate $n=5$ because it more closely traces the highest density regions, yet still shows smaller clumps of sources absent in maps with higher $n$.
The 64 by 64 pixel grid size is sufficient to see small-scale variations in the source density, so a higher resolution map is not necessary.

Our ONC sample spans a greater area, while the OMC1 sample is more centrally concentrated.
Though the ONC source distribution appears to decline towards the edges of our FOV, this is likely at least in part due to the large increase in RMS from the primary beam correction (see Table~\ref{tab:bands}).

\begin{figure}
    \centering
    \includegraphics[width=0.45\textwidth]{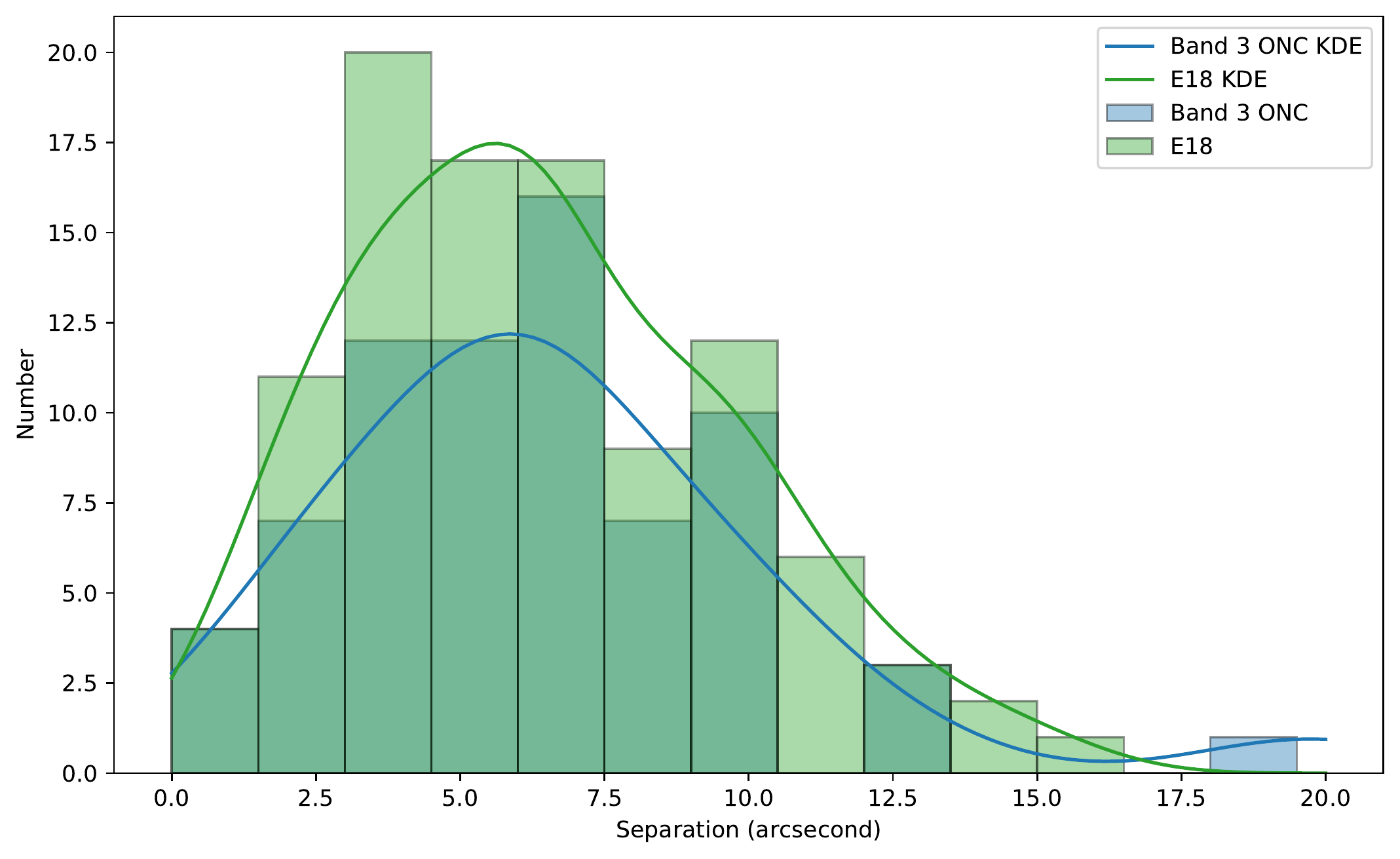}
    \includegraphics[width=0.45\textwidth]{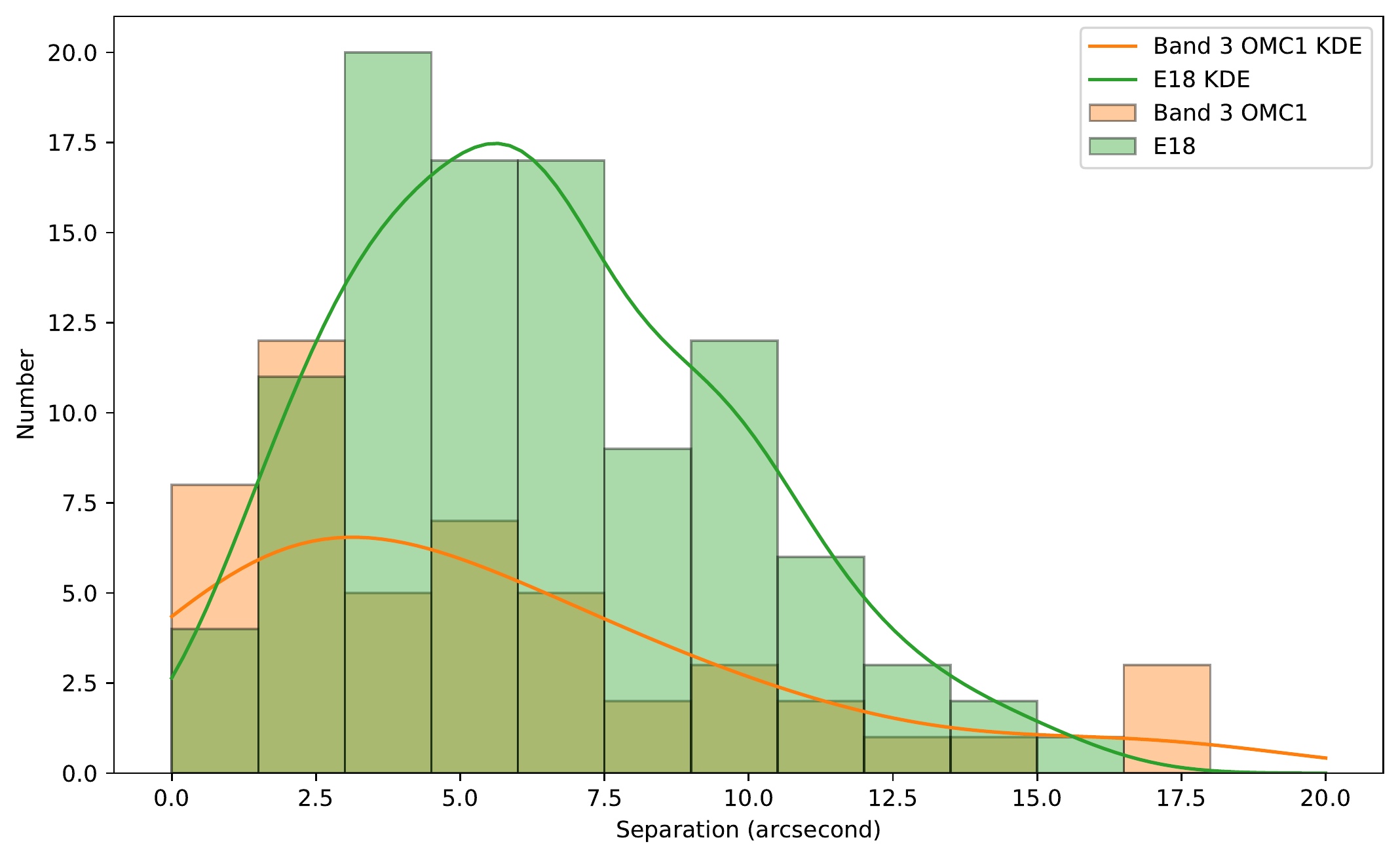}
    \caption{Left: A histogram of source separations in the minimum spanning tree for our band 3 ONC sample (IR detected) and \citetalias{eisner_protoplanetary_2018}. The median separation is $6.2 \pm 0.5$ arcseconds (2480 AU) for our ONC sources, and 5.9 $\pm$  0.3 arcseconds (2360 AU) for \citetalias{eisner_protoplanetary_2018}. Right: The same, but with our band 3 OMC1 sample, with a median separation of 4.6 $\pm$ 0.8 arcseconds (1840 AU).
    In both plots, a Gaussian KDE  with bandwidth of 0.42 (0.46) for the ONC (OMC1) sample and 0.40 for \citetalias{eisner_protoplanetary_2018} is overlaid to better illustrate the distribution shape.}
    \label{fig:sep_hist}
\end{figure}

The OMC1 sample has a higher spatial density than the ONC sample.
We compute the Minimum Spanning tree (MST) of our source locations for our ONC and OMC1 samples and those of \citetalias{eisner_protoplanetary_2018}, then plot a histogram of the source separations in Figure~\ref{fig:sep_hist}.
We use the median separation rather than the mean because the mean OMC1 separation is impacted strongly by sources far from the center whose OMC1 membership is questionable (see Section~\ref{sec:ir_dets}).
The median 2D MST separation for our ONC sample is $6.2 \pm 0.5$ arcseconds, or approximately 2480 AU, for the OMC1 sample it is 4.6 $\pm$ 0.8 arcseconds, or 1840 AU, and for \citetalias{eisner_protoplanetary_2018} it is 5.9 $\pm$ 0.3 arcseconds or 2360 AU, where the errors are the standard deviation of the mean of the MST side lengths.
We do not find any significant ($\gtrsim$ 2$-\sigma$) differences.
In Figure~\ref{fig:sep_hist}, we see that the spatial distribution of our ONC sample is consistent with that of \citetalias{eisner_protoplanetary_2018}. 

In contrast, our OMC1 spatial distribution, shown in Figure~\ref{fig:sep_hist} appears different from that of \citetalias{eisner_protoplanetary_2018}.
A KS-test between the two distributions yields a significant p-value of 0.010.
The OMC1 spatial distribution peaks at around 2 arcseconds while the \citetalias{eisner_protoplanetary_2018} peaks closer to 4 arcseconds, which is reflected in the higher median MST 2D separation.
The higher density of the OMC1 sample is also shown in Figure~\ref{fig:density_3panel} where we see smaller 5th neighbor separations and a more compact spatial distribution than the ONC sample.

The volume density of stars determines the frequency of dynamical interactions, so we need to determine the 3D density from our observed 2D density.
Assuming the source coordinates are distributed as a Gaussian in three dimensions with the same $\sigma$ as observed in two dimensions, we run a simple simulation and find that the median 3D distance to the nearest source is 1.3 times greater than the median 2D distance\footnote{We draw 3D source locations for 10,000 points, compute the median projected 2D and 3D separations for each and compare them. We find similar results for uniform, Gaussian, exponential, and power distributions with the scale factor varying from 1-10.}, hence the typical nearest stellar neighbor distance is $\sim$3000 AU for the OMC1 sample.
Then we approximate the mm source densities as $n_* \sim 3/4\pi d_{3D}^3$, where $d_{3D}$ is the median 3D separation, and find $n_*(\textrm{ONC}) = (6 \pm 1) \times 10^4$ pc$^{-3}$ and $n_*(\textrm{OMC1}) = (16 \pm 3)\times 10^4$ pc$^{-3}$, and $n_*(\textrm{E18}) = (7.3 \pm 0.9)\times 10^4$ pc$^{-3}$ for \citetalias{eisner_protoplanetary_2018}, with errors propagated from above.
Our ONC sample has a similar density to that of \citetalias{eisner_protoplanetary_2018} with similar sensitivities, and is about a factor of three higher than previous central density estimates from \citet{hillenbrand_preliminary_1998}.
Our ONC density estimate only includes mm-detected sources, which are disk-bearing YSOs and some O-stars.
Our ONC source density is a rough estimate given that we do not observe the entire ONC cluster, so our projection corrections could be a severe underestimate.
On the other hand, our OMC1 sample seems to be contained within our field of view, so our projection correction is applicable, though there may be sources with small disks that we do not detect, increasing the stellar density (discussed more in Section~\ref{sec:dmasses}).

The high stellar density inferred in the OMC1 core suggests that it has similar gas and YSO mass.
While we do not know the masses of the YSOs in our sample, we can take $M=0.15$ M$_\odot$ as the median mass of a Kroupa IMF to guess at a stellar mass.
For a density $n_* = \num{1.6e5}$ pc$^{-3}$, the corresponding mass density is $\rho = $\num{1.6e-18} g cm$^{-3}$.
The estimated gas density in the central OMC core is $n(H_2)\lesssim10^6$ cm$^{-3}$, which is $\rho\lesssim\num{4e-18}$ g cm$^{-3}$
\citep[e.g.][]{peng_apex-champ_2012, tang_kinetic_2018}
i.e., it is at most similar to the stellar mass density.
Notably, though, as long as these stars are less than $M<1$ M$_\odot$, they total to $M_{cluster} < M_{SrcI}$, which in the central $\sim10\arcsec\xspace$ that encompasses the inner OMC1 cluster, dominates the mass density with $\rho(4000 AU) = \num{1.4e-16}$ g cm$^{-3}$.

\subsection{Separation from $\theta^1$ Ori C: Does O-star feedback determine disk size \& mass?} \label{sec:theta1c}

\begin{figure}
    \centering
    \includegraphics[width=0.8\textwidth]{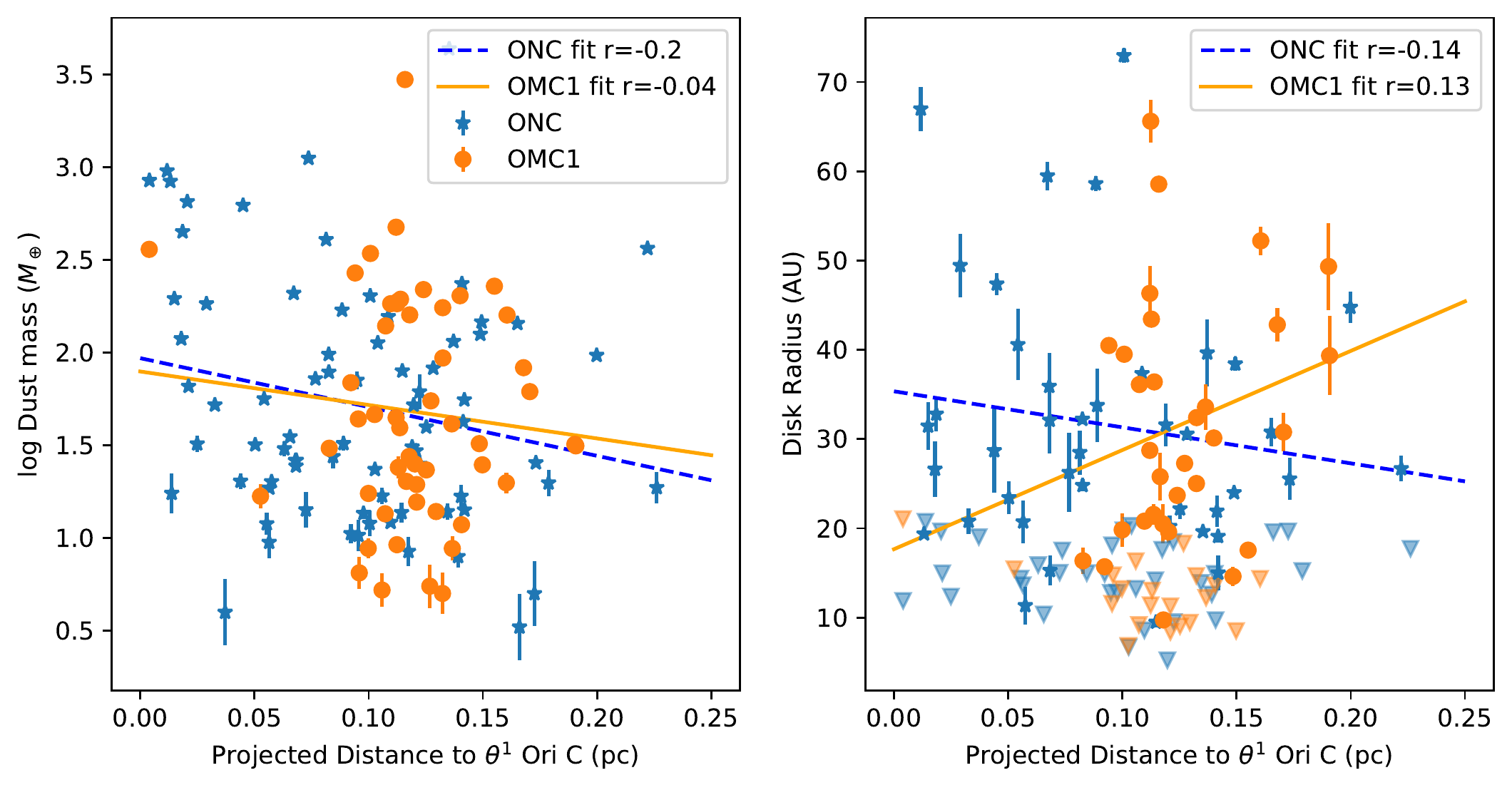}
    \caption{Left: log disk dust mass as a function of projected distance to $\theta^1$ Ori C, for our ONC (blue) and OMC1 (orange) samples, with Spearman correlations of -0.092 and 0.020 for the ONC and OMC1 samples respectively. Right: The same but we plot disk radius, with upper limits shown as triangles. Our ONC (OMC1) sample has a Spearman correlation of -0.090 (0.016).}
    \label{fig:theta1c}
\end{figure}

Feedback from massive stars may result in disk truncation from external photoevaporation.
Naively, then, one may expect that the disk size or mass should decrease with proximity to the massive $\theta^1$ Ori C.
However, \citet{parker_external_2021} show that the combination of projection effects and stellar dynamics serve to mask this effect.
Indeed,  \citetalias{eisner_protoplanetary_2018} find only a marginally significant correlation (\citet{parker_external_2021} find an even weaker correlation in the same data), however \citet{mann_alma_2014}, another sub-mm ONC study, find a very strong correlation, although over a much larger distance, up to $\sim1$ pc. 
We perform the same test with our own deeper, more complete data. 
In Figure~\ref{fig:theta1c}, we plot the disk dust mass and radius of our ONC and OMC1 sample as a function of projected distance to $\theta^1$ Ori C, the nearest massive star and source of ionizing photons.
We linearly fit these data for our ONC and OMC1 samples and find the following relations:
\begin{eqnarray}
    \log (M_{\textrm{dust, ONC}}/M_\oplus) = (-2.6 \pm 1.4)D/\textrm{pc} + (1.94 \pm 0.15) \\
    \log (M_{\textrm{dust, OMC1}}/M_\oplus) = (-1.8 \pm 2.8)D/\textrm{pc} + (1.87 \pm 0.34) \\
    R_{\textrm{disk, ONC}}/\textrm{AU} = (-40 \pm 43)D/\textrm{pc} + (35 \pm 5)  \\
    R_{\textrm{disk, OMC1}}/\textrm{AU} = (110 \pm 90)D/\textrm{pc} + (17 \pm 11)
\end{eqnarray}
We see that all of these fits have slopes consistent with zero within 2$-\sigma$, and that qualitatively these relations are dominated by scatter.
We do not find any significant correlations from a Spearman rank $\rho$ test, with p-values of 0.09 (0.77) and 0.37 (0.48) for ONC (OMC1) dust masses and disk radii respectively.
We note that we do not detect $\theta^1$ Ori C, and find a band 3 flux upper limit of 0.42 mJy; the closest source in this plot is 124 at a separation of 810 $\pm$ 20 AU

\section{ONC/OMC1 in context: Source Populations and Scaling Relations} \label{sec:source_pop}

\subsection{The Luminosity-Radius Relation} \label{sec:scaling_rels}
The disk sample reported here is smaller and less luminous than in other regions. 
We consider the scaling relations between disk size and luminosity by \cite{andrews_scaling_2018} (hereafter \citetalias{andrews_scaling_2018}) in the Lupus star-forming region. 
We scale our flux measurements to their equivalents at the distance to Lupus, i.e., $S_{scaled} = (\SI{400}{pc} / \SI{140}{pc})^2 S_{measured}$, to directly compare to the \citetalias{andrews_scaling_2018} scaling relations.

\begin{figure}
    \centering
    \includegraphics[width=0.8\textwidth]{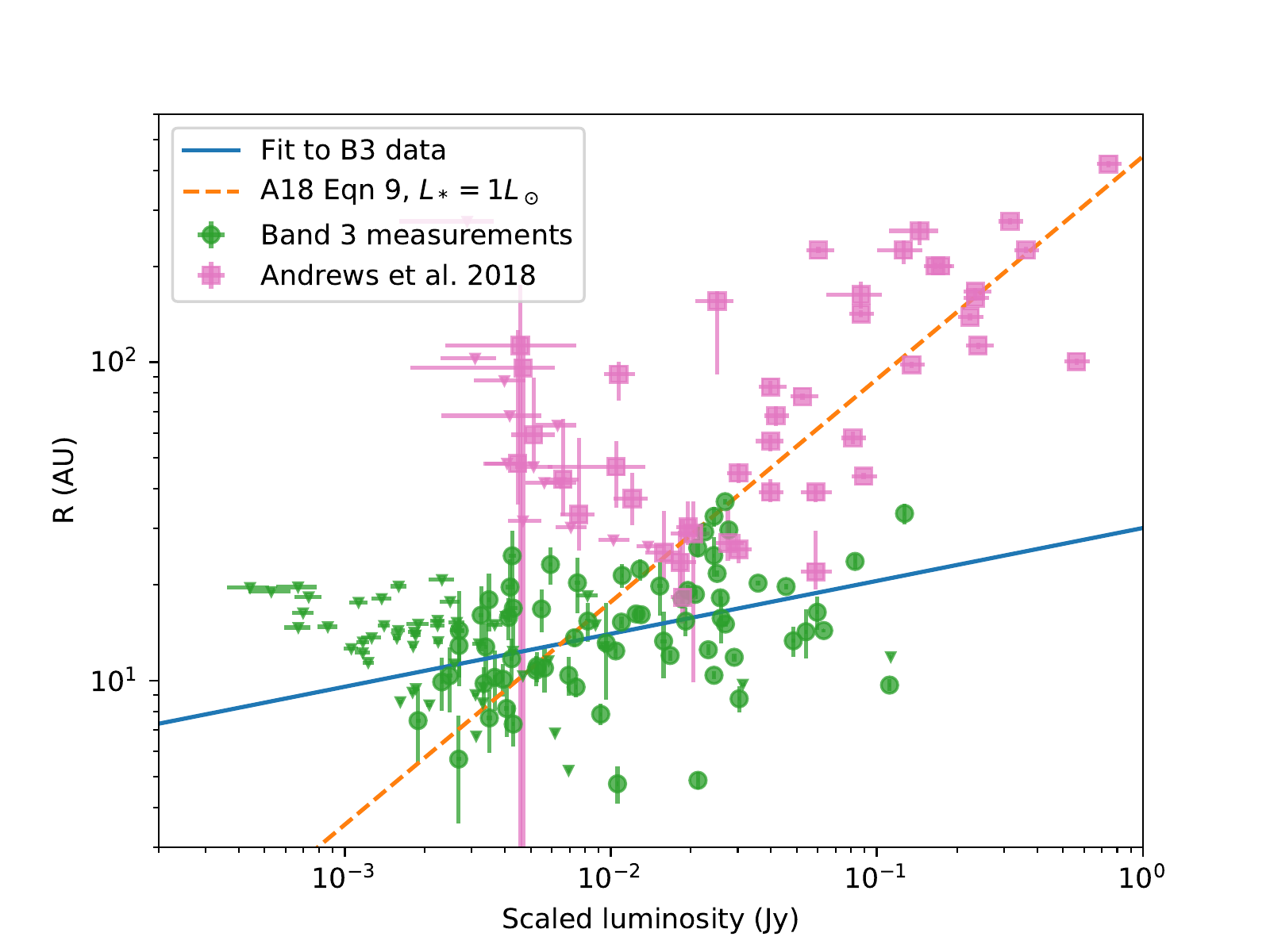}
    \caption{A plot of disk HWHM versus scaled millimeter luminosity. Green points are our band 3 measurements, and pink points are data from \cite{andrews_scaling_2018}.  Triangles represent upper limits. The orange dashed line shows the radius and luminosity scaling relation derived by \citetalias{andrews_scaling_2018} for disks with solar luminosity protostars. The blue solid line is a linear fit to our band 3 measurements, with sources I and BN removed as outliers. This fit has a slope of 0.17 $\pm$ 0.05 AU/Jy and correlation $r^2 = 0.15$.}
    \label{LR_relation}
\end{figure}

We plot the luminosity and disk size as the red dashed line in Figure~\ref{LR_relation}, and plot Equation~8 in \citetalias{andrews_scaling_2018} with $L_* = L_\odot$. 
We exclude Src I and BN (the brightest sources in our data) because they are massive stars that clearly have no equivalents in any of the other observed regions.
We plot a linear least-squares fit of the band 3 and \citetalias{andrews_scaling_2018} data.
This linear fit has equation $R \text{[AU]}= (0.51 \pm 0.05)S_\nu \text{[Jy]} + 2.30 \text{[AU]}$ with $r^2$ = 0.46, which is shallower than that of Equation~8 in \citetalias{andrews_scaling_2018}.
Considering the band 3 data independently, we measure the best-fit equation $R \text{[AU]} = (0.17 \pm 0.05) S_\nu \text{[Jy]} + 1.48 \text{[AU]}$ with $r^2$ = 0.15, and hence find little correlation.
The slope is steeper when including \citetalias{andrews_scaling_2018} than considering the band 3 data alone, with an increase of 4.8-$\sigma$.

While the general trend of \citetalias{andrews_scaling_2018} in Figure~\ref{LR_relation} is preserved when adding our measurements, there is no clear relation in our data alone.
\citetalias{andrews_scaling_2018}  contains disks from the Lupus star-forming region, a young ($\sim$ 1-2 Myr) and nearby ($\sim$150-200 pc) star-forming region.
We see that the disk sizes in Lupus are generally larger than our band 3 sizes in Figure~\ref{LR_relation}.
Our sample lacks disks with HWHM$>$40 AU, where most of the \citetalias{andrews_scaling_2018} sample resides.
Lupus is a less rich star-forming region than Orion, so disk truncation processes that are more effective in the ONC/OMC1 may drive this difference in size.
We discuss disk truncation mechanisms in Section~\ref{sec:discussion} below.
\citet{hendler_evolution_2020} similarly find differing slopes for this scaling relation in different regions.

Observed dust disk sizes are expected to decrease over time even in isolation \citep[e.g.,][]{rosotti_time_2019}.
However, the age of the ONC remains controversial, with potentially substantial age spread and disagreement between different evolutionary tracks \citep[][]{jeffries_no_2011,  messina_impact_2017, winter_solution_2019, fang_improved_2020}.
Within these uncertainties, the age of Orion cannot be distinguished from Lupus, so it is unclear whether the size difference can be attributed to age.

\subsection{Disk Sizes}
\label{sec:disksizes}

\begin{figure}
    \centering
    \includegraphics[width=\textwidth]{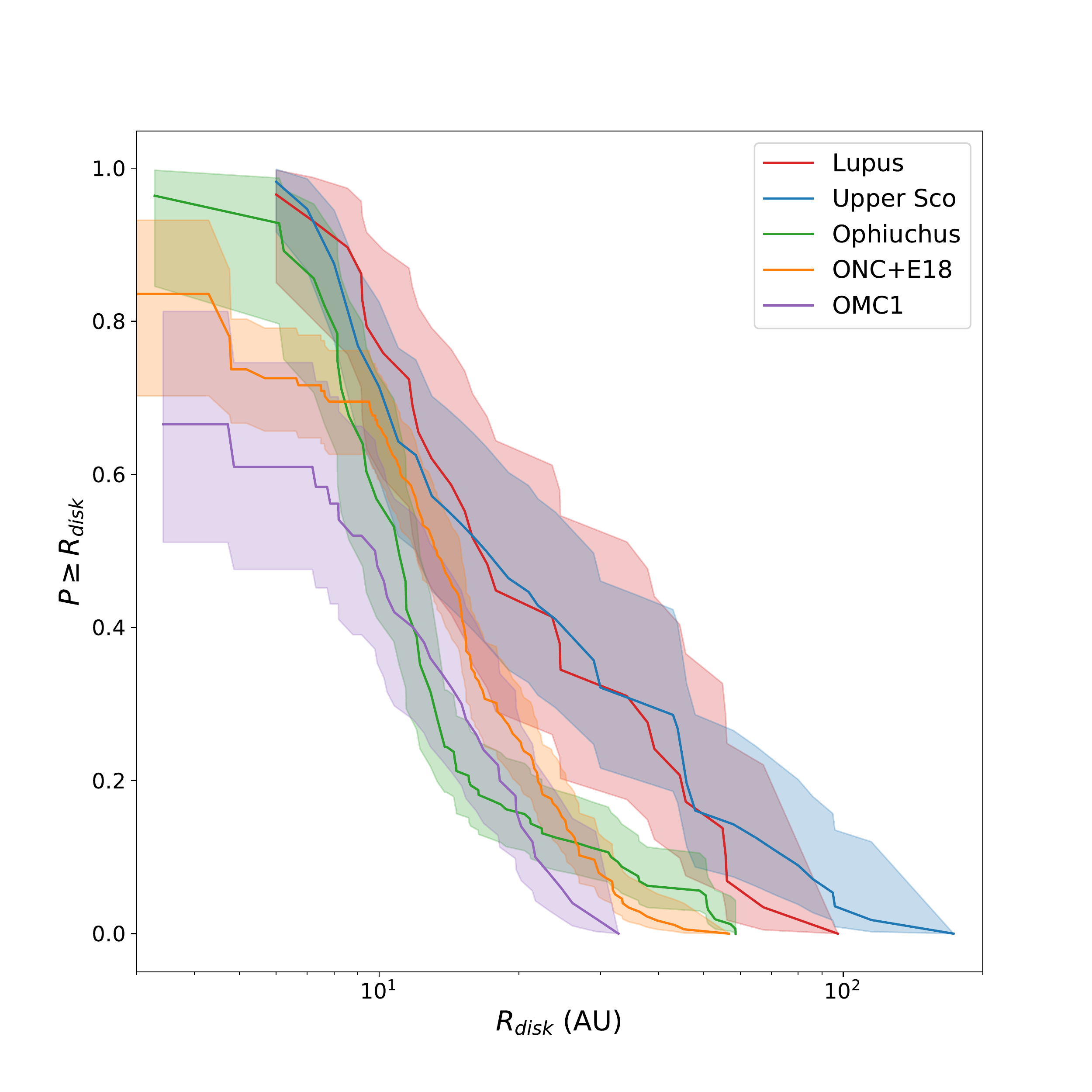}
    \caption{Survival functions of disk radius in Lupus \citep{ansdell_alma_2016}, Upper Scorpius \citep{barenfeld_measurement_2017}, Ophiuchus \citep{cieza_ophiuchus_2019}, Taurus \citep{andrews_circumstellar_2005}, our ONC sample combined with \citetalias{eisner_protoplanetary_2018} (as discussed in Section~\ref{sec:oncvsomc1}), and OMC1 band 3 sample. We plot the half width at half maximum (HWHM) for Ophiuchus, and the ONC and OMC1 samples. The curves do not reach unity because unresolved disks are included as censored data.}
    \label{fig:KM_size}
\end{figure}

Both our Orion samples lack the large ($\gtrsim50$ AU) disks seen in other nearby star-forming regions.
We expand our comparison of disk sizes to other nearby star-forming regions: Lupus \citep{ansdell_alma_2016}, Upper Scorpius \citep{barenfeld_measurement_2017}, and Ophiuchus \citep{cieza_ophiuchus_2019}.
We use the Kaplan-Meier Estimator (calculated with the python package \texttt{lifelines} \citealt{davidson-pilon_camdavidsonpilonlifelines_2020}), to plot the survival functions of disk radii in different regions in Figure~\ref{fig:KM_size} because it includes upper limit disk sizes for unresolved sources.
All the included studies measure disk sizes with a Gaussian except \citet{barenfeld_measurement_2017}, who fit disks with a power-law profile and define a cutoff radius.
This exponential cutoff radius should be similar to the Gaussian half width at half maximum (HWHM) up to a factor of 2 \citep{eisner_resolved_2004}, so Figure~\ref{fig:KM_size} plots the HWHM.
Each study is also conducted with different sensitivities and resolutions, and while this difference may bias the small end of these distributions, it should have a minimal impact on the detection of the largest disks.
The disk sizes in both Orion samples are substantially smaller than those in Upper Sco and Lupus, though they are comparable to those in Ophiucus.

\subsection{Disk Dust Masses}
\label{sec:diskmasses}
While we observe a smaller disk sizes in Orion, disk dust masses in the ONC are similar to other regions while the OMC1 dust mass distribution, though highly uncertain, is at least at slightly higher masses than other regions.

For the same regions as in Section \ref{sec:disksizes} \citep[and Taurus][]{andrews_circumstellar_2005} we compare the disk dust mass distributions. We also use the more complete \citet{williams_ophiuchus_2019} for the Ophiuchus data. 
The disk dust masses in these works are calculated according to Equation~\ref{eq:dmass} except for some sources in \citet{andrews_circumstellar_2005} that were measured with SED fitting.
The assumed dust opacities for each study are similar our adopted value of $\kappa = 2$ cm$^2$/g at 1.3 mm and $\beta = 1$.
The dust temperature of $T_d = 20$ K is also consistent with the exception of \citet{barenfeld_measurement_2017}, who scale the dust temperature with stellar luminosity leading to a range of temperatures of approximately 10 - 30 K.
At 3 mm, this may lead to a difference in the computed dust masses up to a factor of $\sim$4, but on average the masses should be consistent.

\begin{figure}
    \centering
    \includegraphics[width=\textwidth]{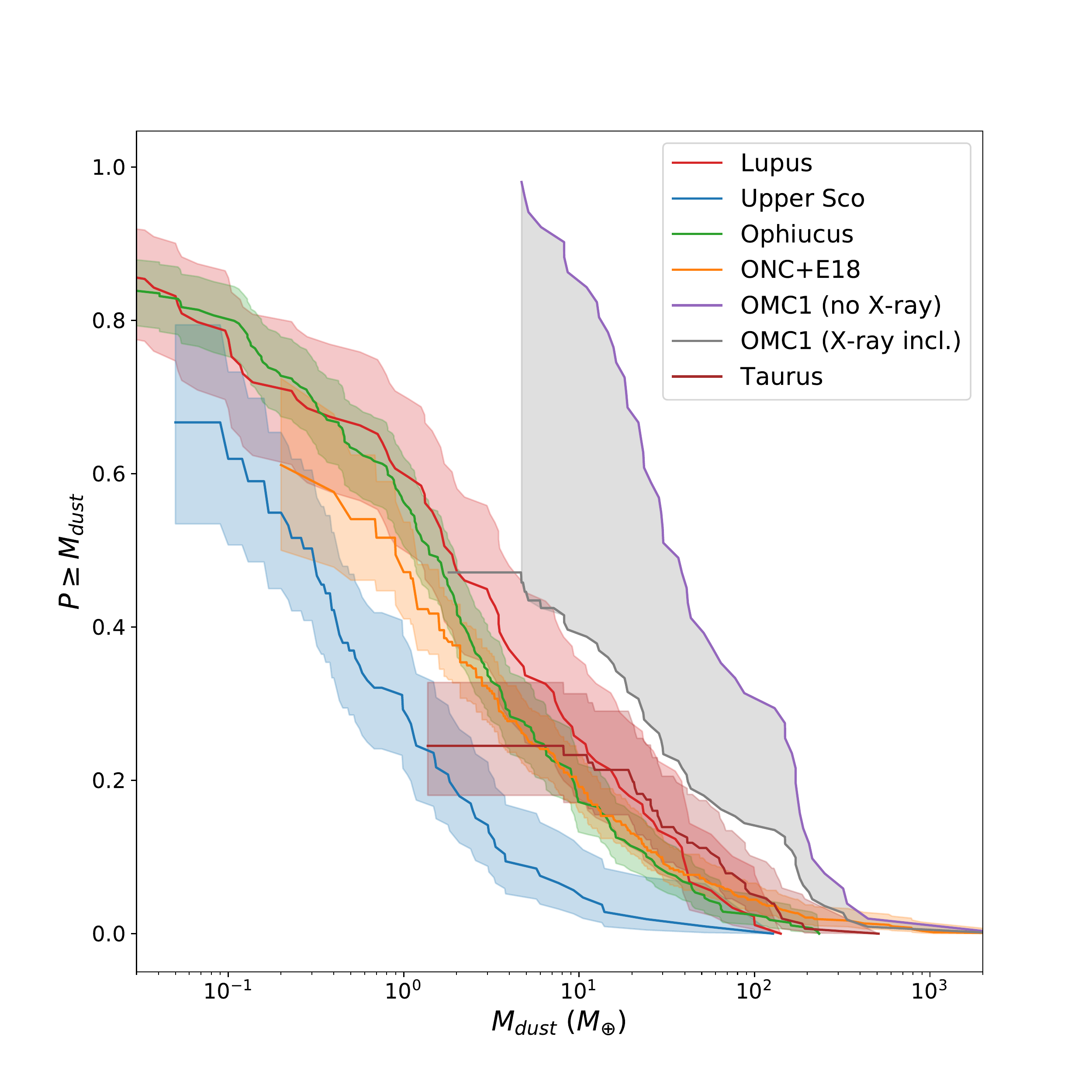}
    \caption{Survival functions of disk dust masses, calculated with Equation~\ref{eq:dmass} in Lupus \citep{ansdell_alma_2016}, Upper Scorpius \citep{barenfeld_alma_2016}, Ophiuchus \citep{cieza_ophiuchus_2019}, Taurus \citep{andrews_circumstellar_2005}, and our ONC and OMC1 samples. The curves do not reach unity because non-detections are included as censored data. There are two curves shown for OMC1, the purple (labeled ``OMC1 (no X-ray)'') only includes disks we detect, and the gray (labeled ``OMC1 (X-ray incl.)'') includes 60 dust mass upper limits from non-detected X-ray sources. The former should be interpreted as an upper limit distribution, and the latter a lower limit (hence the lack of uncertainties).}
    \label{fig:KM_dmass}
\end{figure}

Figure~\ref{fig:KM_dmass} shows the survival functions of disk dust masses in each region and our ONC and OMC1 samples.
To account for the known ONC sources that we do not detect with ALMA, we calculate dust mass upper limits for the 230 non-detected \citetalias{muench_luminosity_2002} sources in our field of view from the band 3 flux upper limits (see Table~\ref{tab:nondet_IR}).
We combine our ONC sample with \citetalias{eisner_protoplanetary_2018} who report similar upper limit fluxes for IR non-detections.
We see that our ONC sample has the a long tail towards high masses due to BN (which has a dust mass of 4060 $\pm$ 10 $M_\oplus$), and has a similar mass distribution to Ophiuchus at low and intermediate masses.

We plot two distributions for OMC1 in Figure~\ref{fig:KM_dmass}, the line in purple shows only detected disks and is effectively an upper limit, while the line in gray includes as censored data the 60 sources detected in other surveys (see below) that we do not detect, and is a lower limit.
Uncertainties are not shown because they are much smaller than the difference between these two distributions.
We identify 18 potential OMC1 X-ray sources from \citet{getman_chandra_2005} that we fail to detect.
We consider X-ray sources in our band 3 field of view that are not IR-detected (determined by a 0.7\arcsec\xspace match to \citetalias{muench_luminosity_2002}, as in Section~\ref{sec:ir_dets}).
Of these, we find 27 sources that we do not detect in our data.
We further require that the non-detected X-ray sources are within 47\arcsec\xspace of Src I, leaving 18 sources remaining.
The chosen radius is the distance of the furthest OMC1 sources we detect (118 and 119) with the exception of source 124, whose OMC1 membership is questionable (see Section~\ref{sec:ir_dets}).
In our OMC1 sample, we have an X-ray detection rate of $\sim30\%$ with 14/51 sources detected.

Given that the OMC1 X-ray sources we do not detect are likely stars with small or negligible disks, we do not expect the X-ray detection rate for OMC1 sources to differ between the mm detected and mm non-detected populations.
Then, we infer that there are 60 total OMC1 sources we fail to detect based on the presence of 18 X-ray OMC1 sources we fail to detect.
Because we only have positions of 18 OMC1 non-detections and need a total of 60 dust mass upper limits, we include the dust mass upper limit for the 18 sources three times each, and randomly select six to be repeated again for a total of 60 upper limits.
We repeat the random selection multiple times and find it has very little impact on the survival function in Figure~\ref{fig:KM_dmass}.

Curiously, none of the X-ray OMC1 non-detections are in the highest density OMC1 region shown in Figure~\ref{fig:density_3panel}, making their OMC1 membership difficult to assess, which is why we treat the distribution including these sources as a lower limit.

Even in the upper limit case, we see the OMC1 sample has higher dust mass disks than other regions.
The large disk dust masses of OMC1 sources and relative lack of low mass disks may be expected because these sources are all embedded in a common envelope, and they therefore have a substantial reservoir to continue accreting from.

\section{Discussion} \label{sec:discussion}

The small sizes of disks in our data, and the low masses in the ONC subsample, as compared with other nearby star-forming clouds suggests that there are additional processes affecting disks in Orion. 
In this section, we discuss disk truncation mechanisms and determine that photoevaporation is a plausible explanation for the small ONC disk sizes, while a combination of dynamical truncation and interactions with the ISM (i.e., ram pressure stripping and face-on accretion) are likely important for the OMC1 population.

While dynamical truncation can significantly decrease disk sizes, it cannot explain the small disks in Orion alone.
\citet{vincke_cluster_2016} model disk dynamical interactions with gas and find that disk sizes are smaller in higher-density clusters, predicting $r\sim50$ AU after 2 Myr at 0.1 pc in a Trapezium-like cluster and $r\sim20$ AU near the center of a $\sim10\times$ denser cluster. 
Our observed disk sizes are $\sim20$ AU in both samples ($r_{\textrm{ONC}}=21\pm4$, $r_{\textrm{OMC1}}=20\pm3$), but the inferred stellar densities are different ($n_*(\textrm{OMC1}) = (16 \pm 3)\times 10^4$ pc$^{-3}$, $n_*(\textrm{ONC}) = (6 \pm 1) \times 10^4$ pc$^{-3}$, $n_*(\textrm{OMC1}) \approx 2.7 n_*(\textrm{ONC})$).
Our observations do not follow the trend in \citet{vincke_cluster_2016} of smaller disks in the denser cluster.
The disk sizes in the joint sample are smaller than predicted by \citet{vincke_cluster_2016} for a Trapezium-like cluster.
These results suggest either that disk truncation is more efficient than they model, and has a weaker dependence on density, or some more efficient truncation process dominates over dynamical truncation.

Disk interactions with the interstellar medium (ISM) during the embedded phase can contribute significantly to decreasing disk sizes.
\citet{wijnen_characterising_2017} model two such interactions: face-on accretion and ram pressure stripping.
The former process contracts disks by adding ISM material without azimuthal angular momentum, thus shrinking the disk to conserve total angular momentum, while the latter removes material due to the pressure exerted by the surrounding gas moving at a different velocity than the disk.
Adopting an OMC1 gas density of approximately $2\times10^{-18}$g cm$^{-3}$ \citep[e.g.][]{peng_apex-champ_2012, tang_kinetic_2018}, in Figure 3 of \citet{wijnen_characterising_2017}, they predict disk radii of about 60$\sim$90 AU with disk velocities ranging from 10 km/s to 1 km/s after a few thousand years.
Their models show that truncation to $\sim20$ AU can occur in a few kyr at densities about an order of magnitude higher; this model matches the observed disk sizes if there is $\gtrsim20$ \msun of gas in the the central $\lesssim6000$ AU of OMC1 (about an order of magnitude higher than observed).

Dynamical truncation may also be important at the high stellar densities observed in OMC1.
\citet{wijnen_disc_2017} model both dynamical and gas-driven disk truncation and conclude that, in environments like OMC1, the two processes are comparable (e.g., their Figure 3a).
Because of the combination of large disk dust masses shown and small disk sizes (shown in Figures~\ref{fig:KM_dmass} and \ref{fig:KM_size} respectively), we suggest the accretion-driven contraction scenario is more likely, as this model suggests that disk masses will increase as their radii decrease.

For our ONC sample, external disk photoevaporation is an efficient truncation mechanism.
\citet{winter_protoplanetary_2018} include external photoevaporation in their simulations and conclude that it is the dominant mechanism of disk truncation for a variety of clusters they simulate, including the ONC.
There is strong observational evidence that disks in the ONC are actively being photoablated  \citep[e.g.][]{mccullough_photoevaporating_1995, bally_externally_1998}.

In Section~\ref{sec:theta1c} we do not find any significant correlations between projected distance to $\theta^1$ Ori C, the nearest source of external ionizing photons, and disk sizes or fluxes.
While previous observational studies have reported correlations between disk properties and projected distance to the nearest ionizing source \citep[e.g.][]{mann_alma_2014, eisner_protoplanetary_2018}, \citet{parker_external_2021} show such correlations can be masked by projection effects and cluster dynamics.

However, it is possible that UV photoevaporation affects only the gas and small grains, not the large grains we observe \citep{haworth_fried_2018,parker_external_2021}.
In this case, external UV radiation may still play a large role in disk truncation, but it will not affect millimeter-observed disk sizes like those presented here.


\section{Summary and Conclusions} \label{sec:conclusions}

We present high resolution band 3, 6 and 7 (3, 1.3, and 0.85 mm) ALMA observations of protoplanetary disks in the ONC and OMC1.
We detect 127 sources in total, with 33 detections in all three bands, and 15 newly detected sources.
We measure spectral indices for 85 sources by including band 7 flux measurements from \citetalias{eisner_protoplanetary_2018}, and find 50/85 sources have spectral indices consistent with $\alpha=2$, indicating many of the sources may be dominated by optically thick dust emission.

We split our sample into two sub-populations, the foreground ONC and the embedded OMC1, and find that the OMC1 population comprises a distinct subcluster from the greater ONC population
We designate the 76 infrared-detected sources as ONC disks, and classify the remaining 51 sources as OMC1 disks. 
We find the OMC1 sample has a higher density than the ONC sample ($n_*(ONC) = (6\pm1) \times 10^4$ pc$^{-3}$ and $n_*(OMC1)=(16\pm3)\times 10^4$ pc$^{-3}$), and that the OMC1 sample is spatially distinct.
Though the disk sizes in the ONC and OMC1 are indistinguishable, OMC1 has an excess of mm-bright disks, hinting at a physical difference between these samples.

We find that the \citetalias{andrews_scaling_2018} scaling relation between millimeter luminosity and disk radius does not seem to describe our resolved sources, though the remaining unresolved sources with upper limit sizes occupy a region potentially consistent with extrapolation from this scaling relation.

The disks in the center of the Orion cluster, in both the ONC and OMC1 samples, are smaller in radius than in other star forming regions, implying that some process is responsible for truncating the disks in this region.
While photoevaporation is a plausible explanation for the ONC subsample, which is exposed to the hard radiation of the Trapezium cluster, it cannot be invoked for the well-shielded OMC1 subsample.
The similarity in disk sizes between these two samples hints instead that dynamical truncation or ISM interactions are important.
We also find that while ONC disk dust masses are consistent with other regions, OMC1 disks tend to have higher dust masses, potentially supporting an accretion-driven truncation scenario.
We conclude that the disparate environments of the ONC and OMC1 effectively truncate disk sizes in both populations.

\begin{acknowledgments}
We thank the anonymous referee for a constructive and detailed review.
JAE acknowledges support from NSF AAG grant 1811290.
A.G. acknowledges support from the National Science Foundation under grant No. 2008101.

\end{acknowledgments}
%

\vspace{5mm}

\facilities{This paper makes use of the following ALMA data: ADS/JAO.ALMA\#2016.1.00165.S ALMA is a partnership of ESO (representing its member states), NSF (USA) and NINS (Japan), together with NRC (Canada), MOST and ASIAA (Taiwan), and KASI (Republic of Korea), in cooperation with the Republic of Chile. The Joint ALMA Observatory is operated by ESO, AUI/NRAO, and NAOJ. The National Radio Astronomy Observatory is a facility of the National Science Foundation operated under cooperative agreement by Associated Universities, Inc.}


\software{This research made use of \texttt{astrodendro}, a Python package to  compute  dendrograms  of  Astronomical  data  (https://dendrograms.readthedocs.io). \texttt{Lifelines} \citep{davidson-pilon_camdavidsonpilonlifelines_2020}, and \texttt{Astropy} \citep{astropy_collaboration_astropy_2013}.}

\appendix

\section{Overview Figures} \label{app:zoom_plots}

\begin{figure}
    \centering
    \includegraphics[width=0.9\textwidth]{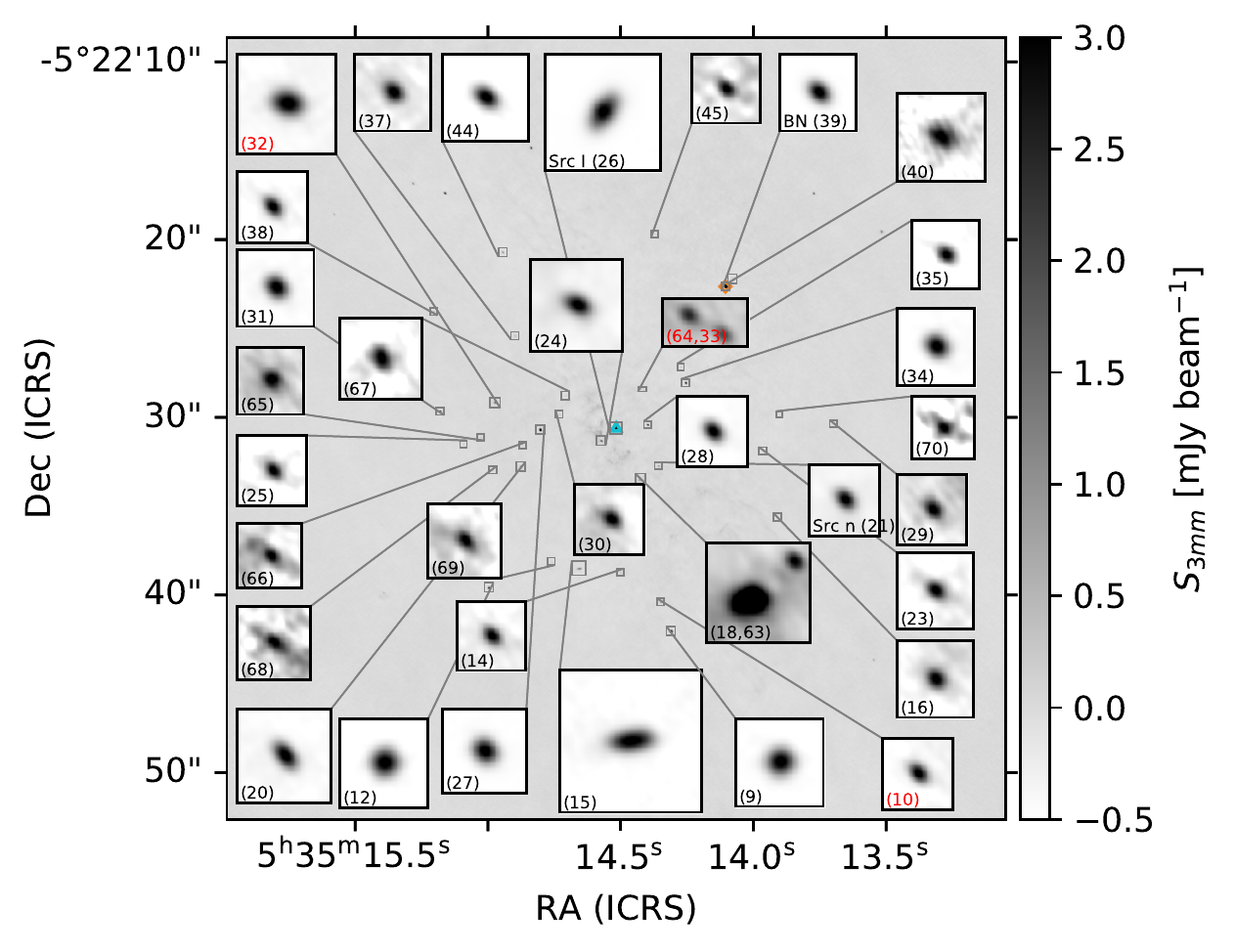}
    \caption{The inner 44\arcsec\xspace of the band 3 (3 mm) image, with the 36 sources detected in the band 7 (0.85 mm) FOV highlighted. Some sources are saturated beyond the 3 mJy colorbar, and others only appear as such with the coloring. Newly detected sources are labelled in red.}
    \label{fig:zoomb7}
\end{figure}

\begin{figure}
    \centering
    \includegraphics[width=0.9\textwidth]{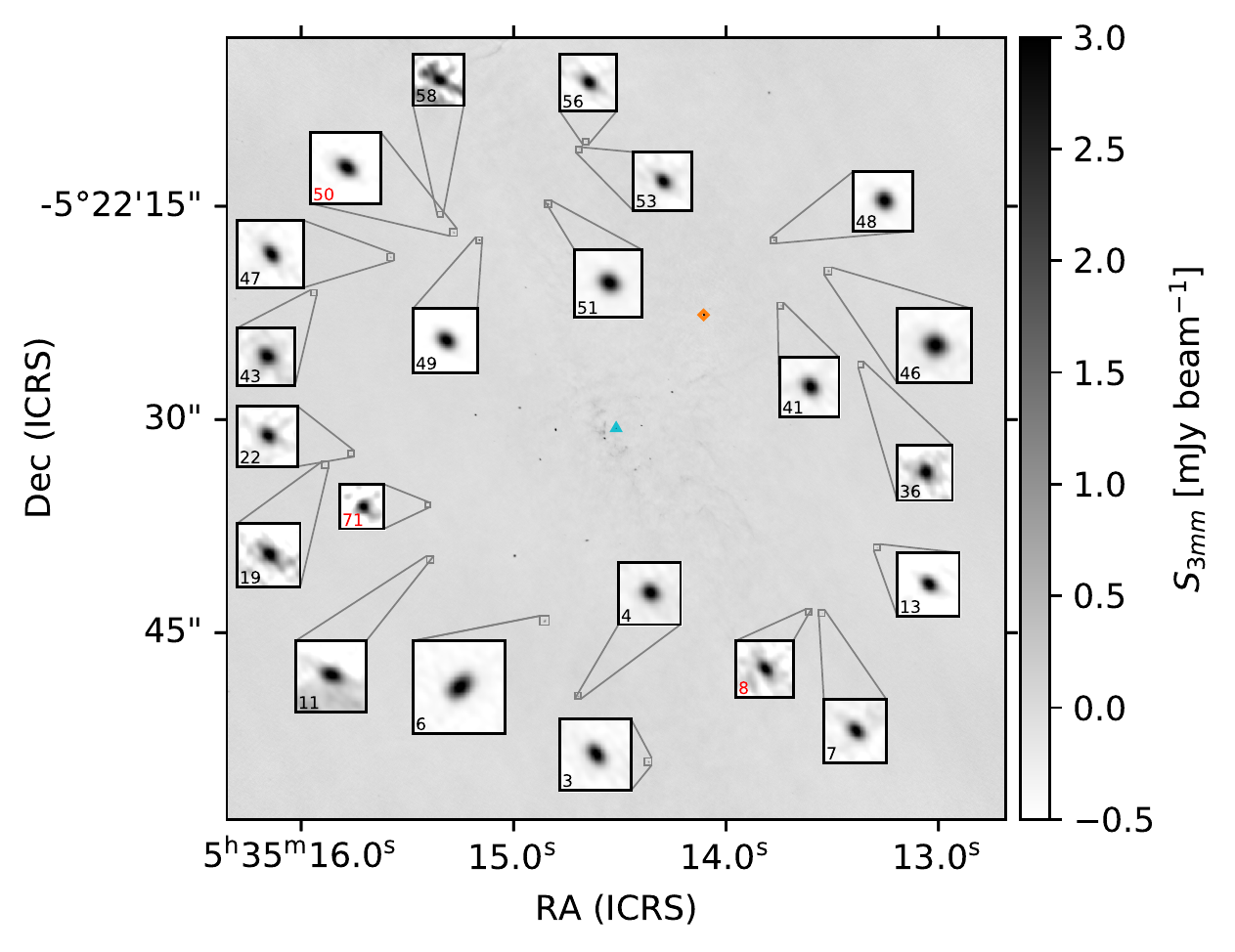}
    \caption{The inner 55\arcsec\xspace of the band 3 (3 mm) image, with the 22 sources detected in the band 6 (1.3 mm) FOV (but outside band 7) highlighted. Some sources are saturated beyond the 3 mJy colorbar, and others only appear as such with the coloring. Newly detected sources are labelled in red.}
    \label{fig:zoomb6}
\end{figure}

\begin{figure}
    \centering
    \includegraphics[width=0.9\textwidth]{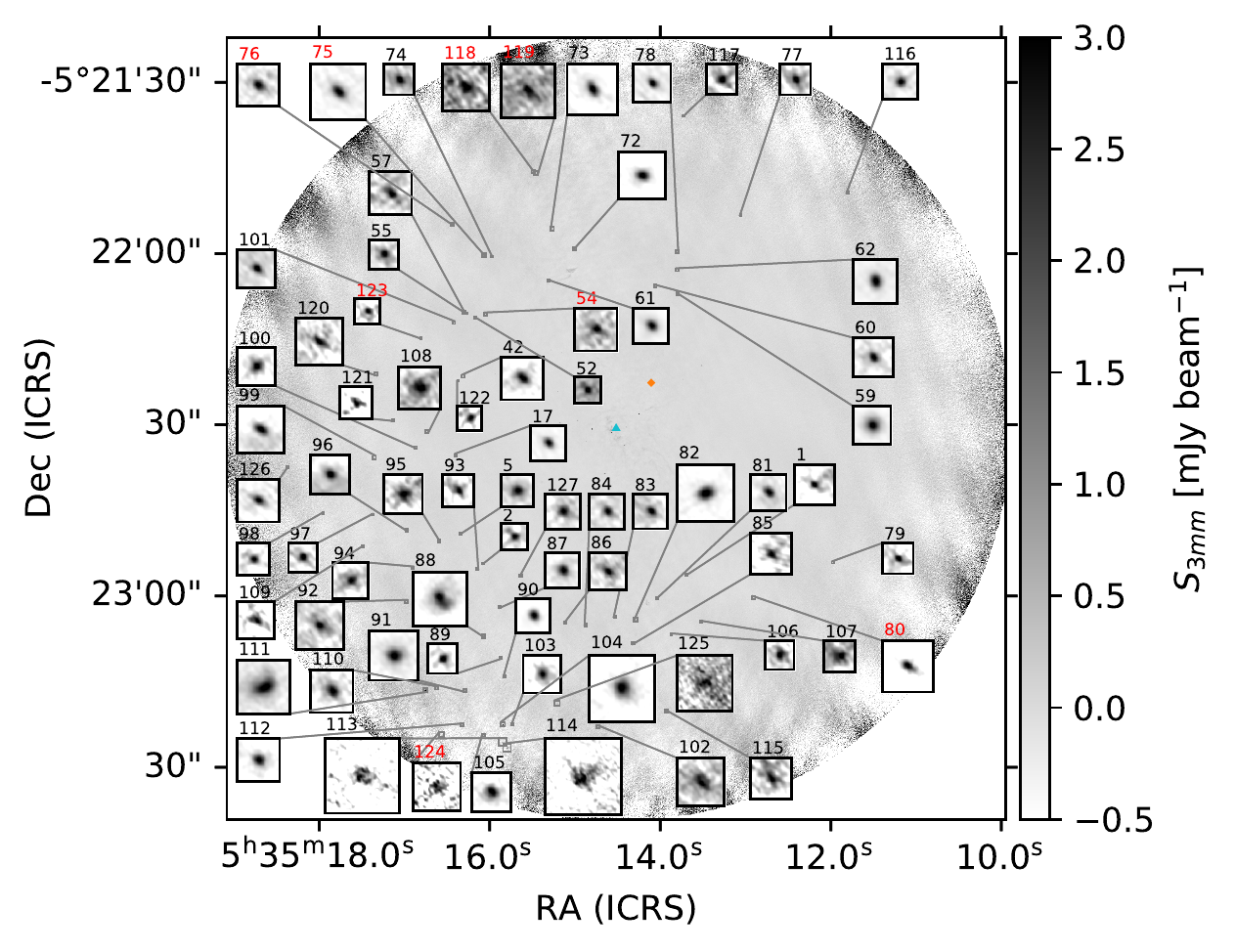}
    \caption{The band 3 (3 mm) image, with the remaining 69 sources detected outside the band 6 and 7 FOVs highlighted. Some sources are saturated beyond the 3 mJy colorbar, and others only appear as such with the coloring. Newly detected sources are labelled in red.}
    \label{fig:zoomb3}
\end{figure}

We include supplemental overview figures of our band 3 data with the locations of each source highlighted.
Figures~\ref{fig:zoomb7}, \ref{fig:zoomb6}, \ref{fig:zoomb3} include sources in the band 7, band 6, and full band 3 FOV respectively.
Sources with red text are newly detected (see Section~\ref{sec:new_dets} and Appendix~\ref{app:newdets}).

\section{Newly detected sources} \label{app:newdets}

For each candidate new source, we list source matches from Vizier and SIMBAD within 2\arcsec\xspace in Table~\ref{tab:new_src}.
We do not include nearby sources if we already determined that they are matched to another source we detect in our catalog (e.g. sources 33 and 64 are within 2\arcsec\xspace of Src I, so we do not include detections of Src I). 
Sources 80, 118, 119, and 123 have no matches within this radius and are thus not included in the table.
Sources 8, 32, 50, 76, and 124 each have matches within 2\arcsec\xspace but the high resolution and astrometric precision of the external studies indicates that these are likely different sources.
The minimum proper motion to associate our detections with the other studies for these sources is 0.11\arcsec\xspace/year, or a projected velocity of $\sim$200 km/s.
This is nearly a factor of ten higher than the velocity of source BN \citep[e.g.][]{rodriguez_proper_2005, gomez_monitoring_2008}, making such a velocity improbable.

Source 71 has a source a mere 1.02\arcsec\xspace away, but the detection is of a peak of $H_2$ line emission corresponding to a clump of dense gas rather than a compact mm source that we observe, as is the case for similar matches to sources 33 and 64 from \citet{nissen_observations_2007}.
The nearby match to source 75 is from a study with Planck data \citep{chen_bandmerged_2016}, which has a resolution larger than our entire field at 5\arcmin.
Sources 10, 33, and 64 are all within 1\arcsec\xspace of a MIR source from \citet{shuping_new_2004}: IRc5 for 10 and IRc21 and IRc2e for 33 and 64.
The mid-infrared (MIR) IRc sources of \citet{shuping_new_2004} are likely dense gas clouds either externally illuminated by massive stars or internally heated by embedded YSOs.
The proximity of our compact sources is consistent with the latter scenario.

\begin{deluxetable}{p{0.03\textwidth}|p{0.15\textwidth}|p{0.09\textwidth}|p{0.06\textwidth}|p{0.09\textwidth}|p{0.08\textwidth}|p{0.08\textwidth}|p{0.06\textwidth}|p{0.08\textwidth}}
\tablecaption{\label{tab:new_src} External sources from Vizier and SIMBAD within 2\arcsec\xspace of our candidate new sources. We do not identify these external sources directly with our sources. (1) ID of matched source in reference. (2) Sky separation from our matched source. (3) Source wavelength or type of emission. (4) Resolution of external study. (5) Quoted astrometric error from external study. (6) Date of observation. (7) Required proper motion to associate the external source to the source in our catalog. (8) Source reference: \citetalias{robberto_wide-field_2010}:\citet{robberto_wide-field_2010}, \citetalias{chambers_pan-starrs1_2016}:\citet{chambers_pan-starrs1_2016}, \citetalias{friedel_high_2011}:\citet{friedel_high_2011}, \citetalias{robberto_hubble_2013}:\citet{robberto_hubble_2013}, \citetalias{sheehan_vla_2016}:\citet{sheehan_vla_2016}, \citetalias{shuping_new_2004}:\citet{shuping_new_2004}, \citetalias{forbrich_population_2016}:\citet{forbrich_population_2016}, \citetalias{feigelson_x-ray-emitting_2002}:\citet{feigelson_x-ray-emitting_2002}, \citet{nissen_observations_2007}, \citetalias{chen_bandmerged_2016}:\citet{chen_bandmerged_2016}, and \citetalias{odell_nature_2015}:\citet{odell_nature_2015}.}
\tabletypesize{\scriptsize}
\startdata
& & & & & & & & \\
ID & External ID$^{(1)}$ & Separation$^{(2)}$ (\arcsec) & Source Type$^{(3)}$ & Resolution$^{(4)}$ & Astrometric error$^{(5)}$ & Epoch$^{(6)}$ & Required proper motion$^{(7)}$ (\arcsec/yr) & Reference$^{(8)}$ \\ \hline
8 & 4272 & 1.89 & NIR & $\sim$0.3\arcsec & 0.15\arcsec & Jan 2005 & 0.15 & \citetalias{robberto_wide-field_2010} \\ 
 & 101540838065555388 & 1.44 & optical & $\sim$0.6\arcsec & 0.03\arcsec & Oct 2012 & 0.29 & \citetalias{chambers_pan-starrs1_2016} \\
 & 101540838062095933 & 1.95 & optical & $\sim$0.6\arcsec & 0.13\arcsec & Dec 2010 & 0.29 & \citetalias{chambers_pan-starrs1_2016} \\ \hline
10 & C28 & 1.59 & mm & 0.5\arcsec & 0.4\arcsec & 2008-2010 & 0.18 & \citetalias{friedel_high_2011} \\ 
 & IRc5 & 0.9 & MIR & 0.38\arcsec & 0.1\arcsec-0.3\arcsec & Nov 2002 & 0.06 & \citetalias{shuping_new_2004} \\ \hline
32 & 4516 & 1.86 & NIR & $\sim$0.3\arcsec & 0.15\arcsec & Jan 2005 & 0.15 & \citetalias{robberto_wide-field_2010} \\
 & 197 (ACS) & 1.76 & optical & - & $<$0.1\arcsec & Oct 2004 & 0.14 & \citetalias{robberto_hubble_2013} \\
 & 4038 (WFPC2) & 1.97 & optical & - & $<$0.1\arcsec & Nov 2004 & 0.15 & \citetalias{robberto_hubble_2013} \\
 & [MLLA] 646 & 1.85 & NIR & 0.5\arcsec-0.6\arcsec & 0.1\arcsec & Mar 2000 & 0.11 & \citetalias{muench_luminosity_2002} \\ \hline
33 & B-11 & 1.08 & $H_2$ line & 0.15\arcsec-0.18\arcsec & - & Dec 2000 & 0.06 & \citetalias{nissen_observations_2007} \\
 & [SEM2016] 306 & 1.6 & radio & 0.1\arcsec & - & Nov 2013 - Mar 2014 & 0.44 & \citetalias{sheehan_vla_2016} \\
 & IRc21 & 0.58 & MIR & 0.38\arcsec & 0.1\arcsec-0.3\arcsec & Nov 2002 & 0.04 & \citetalias{shuping_new_2004} \\
 & IRc2e & 0.83 & MIR & 0.38\arcsec & 0.1\arcsec-0.3\arcsec & Nov 2002 & 0.06 & \citetalias{shuping_new_2004} \\
 & IRc2b & 1.44 & MIR & 0.38\arcsec & 0.1\arcsec-0.3\arcsec & Nov 2002 & 0.1 & \citetalias{shuping_new_2004} \\
 & IRc2d & 1.52 & MIR & 0.38\arcsec & 0.1\arcsec-0.3\arcsec & Nov 2002 & 0.1 & \citetalias{shuping_new_2004} \\ \hline
50 & 227 & 1.6 & radio & 0.3\arcsec & 0.02\arcsec-0.03\arcsec & Oct 2012 & 0.33 & \citetalias{forbrich_population_2016} \\
 & [FBG2002] 469 & 1.97 & Xray &  & 0.25\arcsec & Oct 1999 - Apr 2000 & 0.11 & \citetalias{feigelson_x-ray-emitting_2002} \\
 & 101550838136004995 & 1.47 & optical & $\sim$0.6\arcsec & 0.05" & Mar 2014 & 0.42 & \citetalias{chambers_pan-starrs1_2016} \\ \hline
64 & IRc21 & 0.53 & MIR & 0.38\arcsec & 0.1\arcsec-0.3\arcsec & Nov 2002 & 0.04 & \citetalias{shuping_new_2004} \\
 & IRc2e & 0.85 & MIR & 0.38\arcsec & 0.1\arcsec-0.3\arcsec & Nov 2002 & 0.06 & \citetalias{shuping_new_2004} \\
 & IRc2b & 1.42 & MIR & 0.38\arcsec & 0.1\arcsec-0.3\arcsec & Nov 2002 & 0.1 & \citetalias{shuping_new_2004} \\
 & [SEM2016] 306 & 1.51 & radio & 0.1\arcsec &  & Nov 2013 - Mar 2014 & 0.41 & \citetalias{shuping_new_2004} \\
 & IRc2d & 1.56 & MIR & 0.38\arcsec & 0.1\arcsec-0.3\arcsec & Nov 2002 & 0.11 & \citetalias{shuping_new_2004} \\ 
 & B-11 & 1.29 & $H_2$ line & 0.15\arcsec-0.18\arcsec & - & Dec 2000 & 0.06 & \citetalias{nissen_observations_2007} \\\hline
71 & 2-32 & 1.02 & $H_2$ line & 0.15\arcsec-0.18\arcsec & - & Dec 2000 & 0.06 & \citetalias{nissen_observations_2007} \\ \hline
75 & PLCKERC -070 G208.99-19.37 & 1.05 & FIR & 5\arcmin & - & Aug 2009 - Jan 2012 & 0.15 & \citetalias{chen_bandmerged_2016} \\ \hline
76 & 101560838188622487 & 1.714 & optical & $\sim$0.6\arcsec & 0.02" & Jan 2012 & 0.3 & \citetalias{chambers_pan-starrs1_2016} \\
 & 101560838185142711 & 1.816 & optical & $\sim$0.6\arcsec &  & 2010-2014 & 0.38 & \citetalias{chambers_pan-starrs1_2016} \\ \hline
124 & Theta 1 Ori C & 1.99 & O-star & 0.1\arcsec & - & Nov 2013 - Mar 2014 & 0.44 & \citetalias{sheehan_vla_2016} \\
 & HH 1146 & 1.21 & Herbig-haro object & 0.04\arcsec & - & Jan 2012 & 0.21 & \citetalias{odell_nature_2015} \\
\enddata
\end{deluxetable}

\section{Gaussian Fitting} \label{app:gauss}

For most sources, we have a good Gaussian fit in the band 3 and the convolved band 6 and 7 images. However some of the smallest and faintest sources are not detected in the convolved images, so for these sources we use the original higher resolution images. In band 6, these sources are 30, 36, 45, 56, 63, 68, 70, and 71. In band 7, they are 14, 16, 29, 30, 45, 63, 66, 68, 69, and 70.

\begin{figure}
    \centering
    \includegraphics[width=0.35\textwidth]{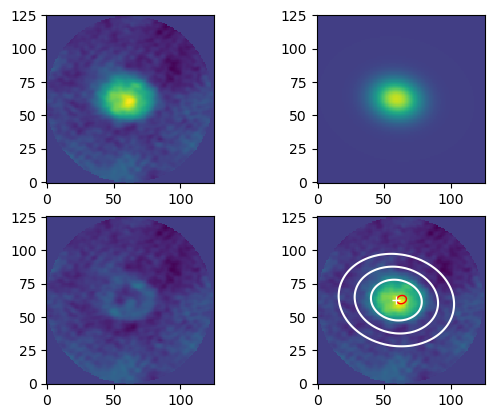}
    \vrule
    \includegraphics[width=0.35\textwidth]{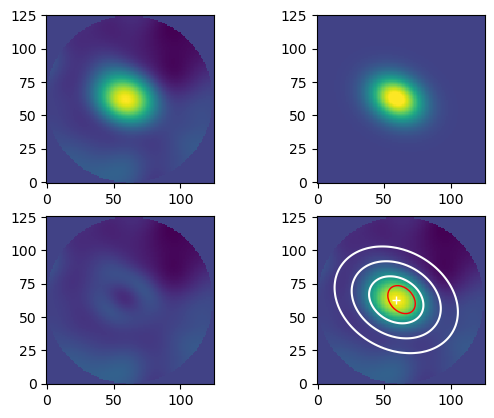}
    \caption{Left: Four panels showing source 32 in the normal band 7 image (top left), the fitted Gaussian (top right), the residuals (bottom left), and contours of the fitted Gaussian in white with the beam in red over the image (bottom right). The axes are in units of pixels. Right: the same four panels except for source 32 in the convolved band 7 image.}
    \label{fig:badfit}
\end{figure}

Figure~\ref{fig:badfit} shows the Gaussian fit for source 32 in both the normal band 7 image (left) and the band 7 image convolved with the band 3 beam (right). There are significant residuals in the fit of the original image (bottom left panel)
The non-convolved fit has a significant ring-like residual and yields a deconvolved size measurement of 0.12 $\pm$ 0.03 arcseconds, while we measure a deconvolved size of 0.097 $\pm$ 0.007 arcseconds with the convolved image.
Thus, we use the convolved band 6 and 7 images for size measurements as long as the source is detected in the convolved images.

\section{Size Upper Limits} \label{app:size}
To deconvolve the beam from the source size, the following conditions must be met.
Let $a_{beam}$, $b_{beam}$, and $\theta_{beam}$ be the major FWHM, minor FWHM, and position angle of the beam, and $a_{src}$, $b_{src}$, and $\theta_{src}$ be the major FWHM, minor FWHM, and position angle fitted Gaussian to the source.
Then we define:
\begin{equation}
    \alpha = (a_{src}\cos(\theta_{src}))^2 + (b_{src}\sin(\theta_{src}))^2 - (a_{beam}\cos(\theta_{beam}))^2 - (b_{beam}\sin(\theta_{beam}))^2
\end{equation}
\begin{equation}
    \beta = (a_{src}\sin(\theta_{src}))^2 + (b_{src}\cos(\theta_{src}))^2 - (a_{beam}\sin(\theta_{beam}))^2 - (b_{beam}\cos(\theta_{beam}))^2
\end{equation}
\begin{equation}
    \gamma = 2((b_{src}^2 - a_{src}^2)\sin(\theta_{src})\cos(\theta_{src}) - (b_{beam}^2 - a_{beam}^2)\sin(\theta_{beam})\cos(\theta_{beam})).
\end{equation}
Then, for a source to be deconvolvable, it must satisfy $\alpha > 0$, $\beta > 0$, and ${\alpha + \beta < \sqrt{(\alpha - \beta)^2 + \gamma^2}}$

\begin{figure}
    \centering
    \includegraphics[width=0.32\textwidth]{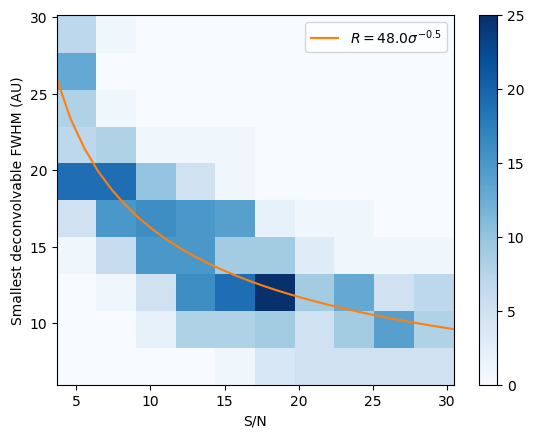}
    \includegraphics[width=0.32\textwidth]{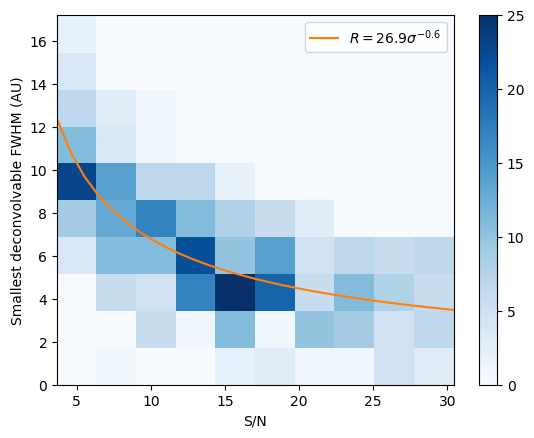}
    \includegraphics[width=0.32\textwidth]{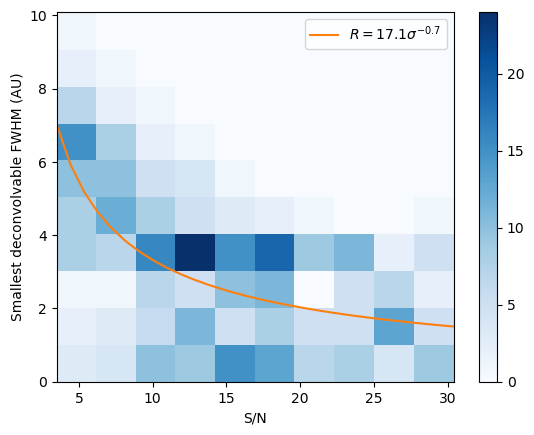}
    \caption{2D histograms showing the relationship between smallest deconvolvable size and simulated disk S/N for left: band 3, middle: band 6, and right: band 7 sources. We fit a power law to each and find $R = 42\sigma^{-0.4}$, $R = 26.1\sigma^{-0.6}$, and $R = 18.3\sigma^{-0.7}$ for bands 3, 6, and 7 respectively.}
    \label{fig:size_ulim}
\end{figure}

We perform the following experiment to determine the conditions needed to obtain a deconvolvable measurement for a given signal-to-noise (S/N) ratio:
We create a synthetic circular disk image and add Gaussian noise such that the beam-scale S/N is in the range 3 to 30.
We convolve this synthetic image with the ALMA beam in B3; this experiment is repeated for B6 and B7.
Starting at 35 AU, we steadily decrease the disk size until the disk can no longer be deconvolved, then record the smallest size at which the disk could be deconvolved, and repeat this process 20 times for 21 different S/N values.
We fit a power law to the smallest deconvolvable size and S/N for simulated disks in each band: $R_{\text{FWHM}} = A(\sigma^B)$, where $R$ is the smallest deconvolvable FWHM in AU, $\sigma$ is the S/N, and $A$ and $B$ are constants.
For sources in each band that cannot be deconvolved, we use this relation to determine the size upper limit given the measured S/N.
For each band, we estimate the following power law relations:
\begin{eqnarray} \label{eq:size_ulim}
    R_{\text{band 3}} & = & 42 \sigma^{-0.43} \\
    R_{\text{band 6}} & = & 26 \sigma^{-0.58} \\
    R_{\text{band 7}} & = & 18 \sigma^{-0.74}
\end{eqnarray}
Where $R$ is the FWHM for a given band and $\sigma$ is the S/N.
Figure~\ref{fig:size_ulim} shows the results of this experiment with 2D histograms of the smallest deconvolvable size as a function of S/N for each band.

\section{Tables} \label{app:tables}

Table~\ref{tab:meas_misc_all} gives an overview of all sources with coordinates and spectral indices for sources detected in multiple bands.
Tables~\ref{tab:meas_misc_all}, \ref{tab:meas_B3}, \ref{tab:meas_B6}, and \ref{tab:meas_B7} give the measured quantities for the 34 sources detected in all bands, where each table is split by band.
These quantities include aperture flux, Gaussian amplitude, measured major and minor axes and position angle, deconvolved major and minor axes and position angle, inclination angle, and calculated dust masses from Equation~\ref{eq:dmass}. 
Note that position angles are measured such that $0^\circ$ corresponds to the major axis lying in the $y$ direction, and increases counterclockwise.

\begin{deluxetable*}{llllrrrrr}
\tablecaption{\label{tab:meas_misc_all} Source coordinates with spectral indices for sources detected in multiple images. $^{(1)}$: spectral index shown in Figure~\ref{fig:alpha_hist}. $^a$: newly detected sources. $^b$: band 6 sources with quantities measured in the original (non-convolved) images. $^c$: band 7 sources with quantities measured in the original images. $^d$: spectral index measured with a band 7 flux from \citetalias{eisner_protoplanetary_2018}. Positional uncertainties do not account for systemic effects from the phase calibration. The spectral index uncertainties include a 20\% systematic flux uncertainty. Only the first six rows are shown.}
\tabletypesize{\scriptsize}
\startdata
& & & & & & \\
ID & MLLA & Forbrich2016 & COUP & Coordinates & $\alpha, \delta$ error & $\alpha_{B3\to B6}$ & $\alpha_{B6\to B7}$ & $\alpha^{(1)}$ \\
1 & 506 &  & 563 & 05:35:13.690 -05:22:56.236 & $\pm 0.005$, $\pm0.004$ & - & - & - \\
2$^d$ & 517 & 274 & 765 & 05:35:16.078 -05:22:54.325 & $\pm 0.004$, $\pm0.003$ & - & - & 1.6$\pm$0.2 \\
3$^d$ & 518 &  & 623 & 05:35:14.366 -05:22:54.054 & $\pm 0.0005$, $\pm0.0006$ & - & - & 1.6$\pm$0.3 \\
4$^d$ & 534 &  & 657 & 05:35:14.697 -05:22:49.426 & $\pm 0.0007$, $\pm0.0007$ & 2.9$\pm$0.3 & - & 1.0$\pm$1.0 \\
5$^d$ & 535 & 296 & 800 & 05:35:16.342 -05:22:49.076 & $\pm 0.002$, $\pm0.002$ & - & - & 1.0$\pm$0.2 \\
6$^d$ & 557 &  &  & 05:35:14.855 -05:22:44.136 & $\pm 0.0008$, $\pm0.0008$ & 2.8$\pm$0.3 & - & 1.0$\pm$1.0 \\
\enddata
\tablecomments{Table~\ref{tab:meas_misc_all} is published in its entirety in the machine-readable format. A portion is shown here for guidance regarding its form and content.}
\end{deluxetable*}

\begin{splitdeluxetable*}{lrrrrrBrrrrrr}
\tablecaption{\label{tab:meas_B3}Measured band 3 quantities for all sources. $^*$: sources poorly fit with a Gaussian with resolved sizes. Their sizes are reported here, but not used in the analysis.
Flux uncertainties do not account for systematic effects from the flux calibration. All sizes are in units of arcseconds. Only the first six rows are shown.}
\tabletypesize{\scriptsize}
\startdata
ID & $F_{\text{ap}}$ & $A_{gaussian}$ & $\text{FWHM}_{\text{maj}}$ & $\text{FWHM}_{\text{min}}$ & $\theta$ & $\text{FWHM}_{\text{maj, deconv}}$ & $\text{FWHM}_{\text{min, deconv}}$ & $\theta_{\text{deconv}}$ & $i$ & $M_{dust, B3}$ \\
1 & 0.17$\pm$0.01 & 0.17$\pm$0.02 & 0.12$\pm$0.01 & 0.065$\pm$0.008 & 58.7$\pm$0.1 & $<0.05$ & - & - & - & 9.7$\pm$0.8 \\
2 & 0.15$\pm$0.01 & 0.23$\pm$0.03 & 0.076$\pm$0.009 & 0.064$\pm$0.007 & 241.7$\pm$0.5 & $<0.03$ & - & - & - & 8.9$\pm$0.8 \\
3 & 1.593$\pm$0.006 & 1.16$\pm$0.01 & 0.125$\pm$0.001 & 0.085$\pm$0.001 & 35.62$\pm$0.02 & 0.081$\pm$0.001 & 0.046$\pm$0.001 & 0.49$\pm$0.02 & 56$\pm$1 & 91.4$\pm$0.3 \\
4 & 1.278$\pm$0.008 & 0.98$\pm$0.02 & 0.109$\pm$0.002 & 0.093$\pm$0.001 & 52.16$\pm$0.07 & 0.062$\pm$0.002 & 0.048$\pm$0.001 & -1.15$\pm$0.07 & 39$\pm$3 & 73.3$\pm$0.5 \\
5 & 0.57$\pm$0.01 & 0.43$\pm$0.02 & 0.094$\pm$0.004 & 0.08$\pm$0.004 & 266.6$\pm$0.2 & $<0.03$ & - & - & - & 32.8$\pm$0.7 \\
6 & 2.76$\pm$0.003 & 1.09$\pm$0.01 & 0.163$\pm$0.002 & 0.114$\pm$0.002 & -46.84$\pm$0.03 & 0.147$\pm$0.002 & 0.061$\pm$0.002 & -46.79$\pm$0.03 & 65.3$\pm$0.8 & 158.3$\pm$0.2 \\
\enddata
\tablecomments{Table~\ref{tab:meas_B3} is published in its entirety in the machine-readable format. A portion is shown here for guidance regarding its form and content.}
\end{splitdeluxetable*}

\begin{splitdeluxetable*}{lrrrrrBrrrrrr}
\tablecaption{\label{tab:meas_B6}Measured band 6 quantities. $^b$: band 6 sources with quantities measured in the original images. Flux uncertainties do not account for systematic effects from the flux calibration. All sizes are in units of arcseconds. Only the first six rows are shown.}
\tabletypesize{\scriptsize}
\startdata
ID & $F_{\text{ap}}$ & $A_{gaussian}$ & $\text{FWHM}_{\text{maj}}$ & $\text{FWHM}_{\text{min}}$ & $\theta$ & $\text{FWHM}_{\text{maj, deconv}}$ & $\text{FWHM}_{\text{min, deconv}}$ & $\theta_{\text{deconv}}$ & $i$ & $M_{dust, B6}$ \\
3 & $<$2840.0 & - & - & - & - & - & - & - & - & $<$8140.0 \\
4 & 14.16$\pm$0.08 & 1.8$\pm$0.2 & 0.11$\pm$0.01 & 0.081$\pm$0.008 & 68.3$\pm$0.3 & $<0.02$ & - & - & - & 80.3$\pm$0.4 \\
6 & 26.97$\pm$0.05 & 2.4$\pm$0.2 & 0.15$\pm$0.01 & 0.12$\pm$0.01 & 262.5$\pm$0.2 & 0.13$\pm$0.01 & 0.084$\pm$0.01 & -83.1$\pm$0.2 & 48$\pm$8 & 153$\pm$0.3 \\
7 & 4.23$\pm$0.05 & 0.8$\pm$0.2 & 0.1$\pm$0.02 & 0.07$\pm$0.01 & 262.3$\pm$0.3 & $<0.03$ & - & - & - & 24$\pm$0.3 \\
8 & 2.92$\pm$0.03 & 0.42$\pm$0.08 & 0.11$\pm$0.02 & 0.08$\pm$0.02 & -3.2$\pm$0.4 & $<0.03$ & - & - & - & 16.6$\pm$0.2 \\
9 & 48.41$\pm$0.02 & 4.5$\pm$0.1 & 0.15$\pm$0.003 & 0.138$\pm$0.003 & -89.7$\pm$0.2 & 0.128$\pm$0.003 & 0.104$\pm$0.003 & -65.3$\pm$0.2 & 36$\pm$3 & 274.7$\pm$0.1 \\
\enddata
\tablecomments{Table~\ref{tab:meas_B6} is published in its entirety in the machine-readable format. A portion is shown here for guidance regarding its form and content.}
\end{splitdeluxetable*}

\begin{splitdeluxetable*}{lrrrrrBrrrrrr}
\tablecaption{\label{tab:meas_B7}Measured band 7 quantities. $^c$: band 7 sources with quantities measured in the original images. Flux uncertainties do not account for systematic effects from the flux calibration. All sizes are in units of arcseconds. Only the first six rows are shown.}
\tabletypesize{\scriptsize}
\startdata
ID & $F_{\text{ap}}$ & $A_{gaussian}$ & $\text{FWHM}_{\text{maj}}$ & $\text{FWHM}_{\text{min}}$ & $\theta$ & $\text{FWHM}_{\text{maj, deconv}}$ & $\text{FWHM}_{\text{min, deconv}}$ & $\theta_{\text{deconv}}$ & $i$ & $M_{dust, B7}$ \\
9 & 57.7$\pm$0.01 & 2.6$\pm$0.2 & 0.122$\pm$0.008 & 0.121$\pm$0.008 & -83.0$\pm$4.0 & 0.1$\pm$0.008 & 0.076$\pm$0.008 & -49.0$\pm$4.0 & 41$\pm$9 & 108.82$\pm$0.03 \\
10 & 7.81$\pm$0.01 & 0.82$\pm$0.06 & 0.107$\pm$0.008 & 0.061$\pm$0.005 & 52.31$\pm$0.09 & $<0.007$ & - & - & - & 14.74$\pm$0.03 \\
12 & 50.39$\pm$0.03 & 2.5$\pm$0.2 & 0.14$\pm$0.01 & 0.107$\pm$0.008 & 267.3$\pm$0.2 & 0.11$\pm$0.01 & 0.064$\pm$0.008 & -78.7$\pm$0.2 & 56$\pm$6 & 95.03$\pm$0.06 \\
14$^c$ & 0.2$\pm$0.07 & 0.4$\pm$0.2 & 0.03$\pm$0.02 & 0.018$\pm$0.008 & -59.8$\pm$0.6 & $<0.02$ & - & - & - & 0.4$\pm$0.1 \\
15 & 34.057$\pm$0.007 & 1.73$\pm$0.05 & 0.23$\pm$0.007 & 0.09$\pm$0.003 & -83.73$\pm$0.02 & 0.216$\pm$0.007 & 0.025$\pm$0.003 & -81.27$\pm$0.02 & 83.4$\pm$0.8 & 64.22$\pm$0.01 \\
16$^c$ & 5.39$\pm$0.05 & 2.2$\pm$0.3 & 0.055$\pm$0.008 & 0.035$\pm$0.005 & 179.1$\pm$0.2 & 0.049$\pm$0.008 & 0.02$\pm$0.005 & 0.0$\pm$0.2 & 66$\pm$8 & 10.17$\pm$0.09 \\
\enddata
\tablecomments{Table~\ref{tab:meas_B7} is published in its entirety in the machine-readable format. A portion is shown here for guidance regarding its form and content.}
\end{splitdeluxetable*}

\bibliographystyle{aasjournal}
\bibliography{aug21.bib}

\end{document}